\documentclass[11pt]{article}

\usepackage[final]{acl}
\usepackage{flushend}
\usepackage[a-1b]{pdfx}
\usepackage{balance}
\usepackage{color}
\usepackage{amsthm}
\usepackage{tcolorbox}
\usepackage[labelfont=bf]{caption}
\usepackage{subcaption}
\usepackage{hyperref}
\usepackage{makecell}
\usepackage{fixltx2e}
\usepackage{rotating,multirow}
\usepackage{etoolbox,xspace}
\usepackage{algorithmicx}
\usepackage{algorithm}
\usepackage{fontawesome}
\usepackage{algpseudocode}
\usepackage{wrapfig}
\usepackage{scalerel}
\usepackage{booktabs}
\usepackage{multirow}
\usepackage{siunitx}
\usepackage{amsmath}
\usepackage{colortbl}   
\usepackage{todonotes}
\usepackage{pgfplots}
\usepackage{marginnote}
\usepackage[dvipsnames]{xcolor}
\usepackage{pifont}
\algtext*{EndWhile}
\algtext*{EndIf}
\algtext*{EndFunction}
\algtext*{EndFor}
\algrenewcommand\algorithmicrequire{\textbf{Input:}}
\algrenewcommand\algorithmicensure{\textbf{Output:}}

\usepackage{amsthm}
\usepackage{tabularx}
\usepackage{ragged2e}  
\usepackage{booktabs}
\usepackage{textcomp}
\usepackage{url}
\usepackage{comment}
\usepackage{enumitem}
\usepackage[simplified]{pgf-umlcd}
\usepackage{arydshln}
\usepackage{tikz}
\usepackage{xspace}
\usetikzlibrary{shapes.geometric, arrows.meta, positioning, calc}

\newcolumntype{Y}{>{\RaggedRight\arraybackslash}X}

\newenvironment{proof}{\paragraph{\sc Proof.}}{\hfill$\square$}
\newtcolorbox{defbox}{
  colback=orange!10,
  colframe=orange!20,
  arc=2mm,
  fonttitle=\bfseries,
  boxrule=0mm,
  boxsep=1mm,
  left=0mm,
  right=0mm,
  top=0mm,
  bottom=0mm
}
 
\newcounter{example}
\renewcommand{\theexample}{\arabic{example}}
\usepackage{tcolorbox}
\newtcolorbox{examplebox}{
  colback=blue!10,
  colframe=blue!20,
  arc=2mm,
  fonttitle=\bfseries,
  boxrule=0mm,
  boxsep=1mm,
  left=0mm,
  right=0mm,
  top=0mm,
  bottom=0mm
}

\newcommand{\squishlist}{
 \begin{list}{$\bullet$}
  { \setlength{\itemsep}{0pt}
     \setlength{\parsep}{1pt}
     \setlength{\topsep}{1pt}
     \setlength{\partopsep}{0pt}
     \setlength{\leftmargin}{1em}
     \setlength{\labelwidth}{1em}
     \setlength{\labelsep}{0.5em} } }
\newcommand{\squishend}{
  \end{list}
}

\newcommand{\stitle}[1]{\vspace{2mm}\noindent{\bf #1:}\xspace}

\definecolor{low}{rgb}{1, 0.8, 0.8}   
\definecolor{mid}{rgb}{1, 1, 0.8}     
\definecolor{high}{rgb}{0.8, 1, 0.8}  
\definecolor{DarkGreen}{rgb}{0.0, 0.8, 0.0}
\definecolor{americanrose}{rgb}{1.0, 0.01, 0.24}
\definecolor{airforceblue}{rgb}{0.36, 0.54, 0.66}
\definecolor{ao(english)}{rgb}{0.0, 0.5, 0.0}
\definecolor{ao}{rgb}{0.0, 0.0, 1.0}

\newcommand{\Schema}{\mathcal{S}}           
\newcommand{\Relation}{\mathcal{T}}         

\newcommand{\Database}{\mathcal{D}}         
\newcommand{\Instance}{R}

\newcommand{\AttrRel}{\mathcal{A}_i}        
\newcommand{\AttrS}{\mathcal{A}^S}          
\newcommand{\AttrU}{\mathcal{A}^U}          
\newcommand{\Tuple}{r}

\newcommand{\Question}{\mathcal{Q}}

\newcommand{\System}{\textsc{OmniTQA}\xspace}
\newcommand{\Preview}{\hat{R}^\Question}
\renewcommand{\arraystretch}{1.3}

\usepackage{listings}
\usetikzlibrary{shapes.geometric,arrows} 

\tcbuselibrary{breakable, listings}
\usepackage{cuted}

\usepackage{times}
\usepackage{amssymb}

\usepackage{latexsym}

\usepackage[T1]{fontenc}

\usepackage[utf8]{inputenc}

\usepackage{microtype}

\usepackage{inconsolata}

\usepackage{graphicx}

\usepackage{xcolor}

\newcommand{\seiji}[1]{{#1}}

\definecolor{deepgreen}{rgb}{0.0, 0.5, 0.0}
\definecolor{sigdrop}{rgb}{1.0, 0.5, 0.5}
\definecolor{moddrop}{rgb}{1.0, 0.8, 0.8}
\definecolor{slidrop}{rgb}{1.0, 0.9, 0.9}

\newcommand{\xmark}{\textcolor{red}{\textbf{\ding{55}}}}
\newcommand{\cmark}{\textcolor{deepgreen}{\textbf{\ding{51}}}}

\title{From Textual Columns to Query Plans: A Unified Relational-Semantic Execution Framework for Hybrid Query Processing}

\author{
  Nima Shahbazi \\
  Megagon Labs \\
  \texttt{nima@megagon.ai} \\
  \And
  Seiji Maekawa \\
  Megagon Labs \\
  \texttt{seiji@megagon.ai} \\
  \And
  Nikita Bhutani \\
  Megagon Labs \\
  \texttt{nikita@megagon.ai} \\
  \And
  Estevam Hruschka \\
  Megagon Labs \\
  \texttt{estevam@megagon.ai} \\
}

\begin{document}
\maketitle
\begin{abstract}
Real-world table question answering often involves hybrid schemas in which some query-relevant information is explicit in relational columns, while other attributes, predicates, or join conditions are only implicit in free-form text. Existing systems struggle with this setting: Text-to-SQL methods scale to large and multi-table databases but require fully structured schemas, whereas direct LLM-based methods can interpret textual content but are costly and unreliable when applied to large databases. We present \System, a unified framework for semi-structured table question answering that treats semantic reasoning as a first-class operation within relational query execution. \System compiles natural-language questions into directed acyclic graphs of relational and LLM-based semantic operators. This enables ambiguity-aware plan diversification, cost-aware optimization, and dual-engine execution over structured and textual data. Across structured and semi-structured benchmarks, \System consistently improves performance in hybrid settings, outperforming the strongest baselines by 14 accuracy points on average and by 27 points on the most challenging subset, while maintaining competitive accuracy on fully structured datasets.

\end{abstract}

\section{Introduction}
\label{sec:introduction}

\begin{figure}[t]
    
    \centering  \includegraphics[width=\linewidth]{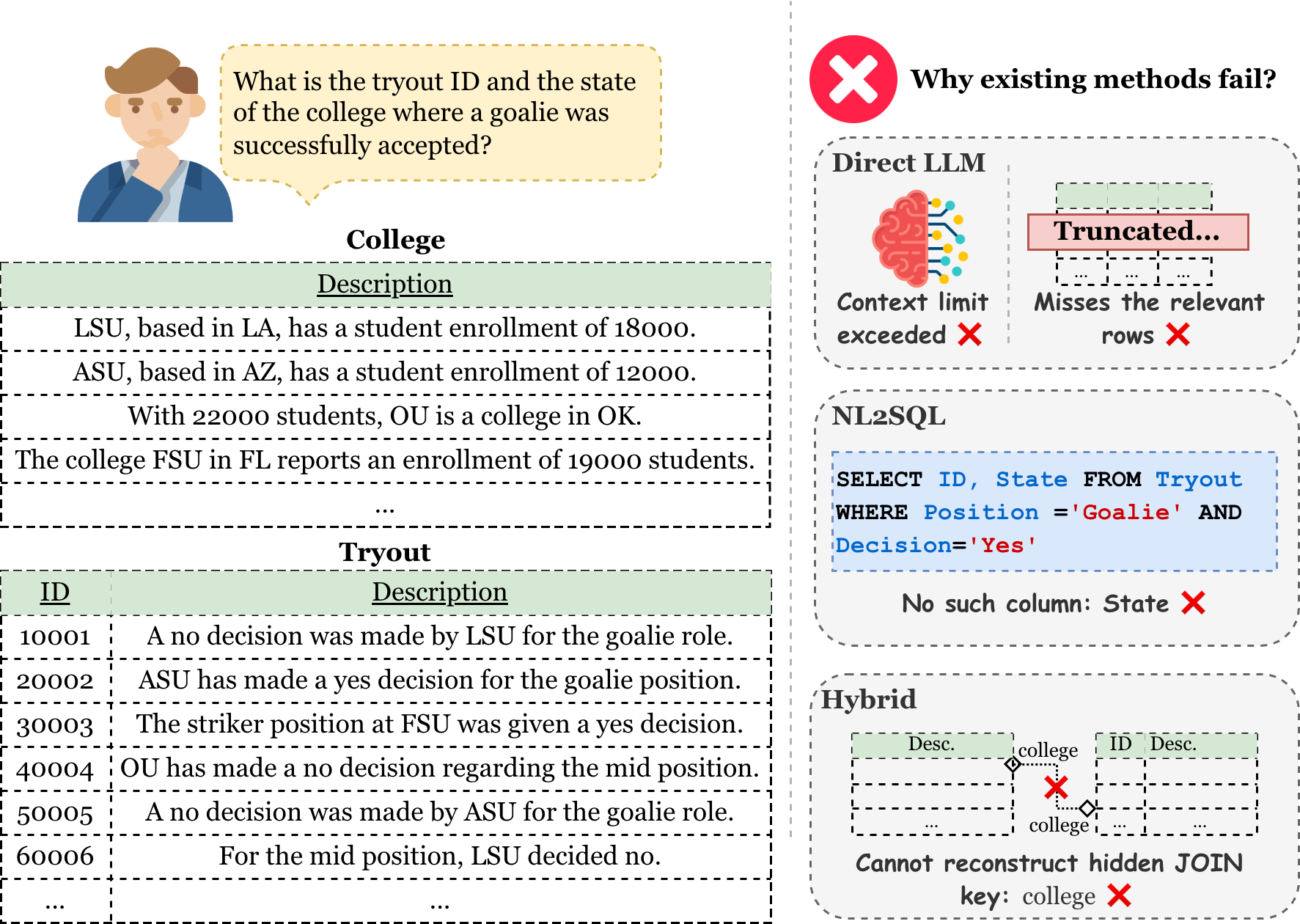}
    \caption{Limitations of existing TQA approaches on querying semi-structured data}
    \label{fig:example-figure}
    
\end{figure}

Recent advances in large language models (LLMs) have significantly improved table question answering (TQA) systems \cite{ipeirotis2025natural, liu2025survey, zhang2024natural}. However, existing approaches operate under a polarized trade-off: they either scale efficiently over large databases by assuming fully structured schemas, or provide semantic flexibility over unstructured text at the cost of poor scalability. This assumption rarely holds in real-world settings, where datasets frequently feature \textit{hybrid} schemas that blend structured attributes with free-form text such as product descriptions, customer feedback, or clinical notes. Answering even simple queries in such settings may require recovering latent attributes from text, grounding them to implicit schema elements, and performing multi-step reasoning jointly over structured and unstructured data (Figure~\ref{fig:example-figure}).

Existing paradigms fail to natively support the intersection of structural scaling and semantic depth required for hybrid schemas. Traditional Text-to-SQL approaches scale well over multi-table databases but assume a strict closed world and fail when critical information is embedded within free-form text \cite{zhou2025table}. Conversely, direct LLM-based approaches \cite{fanglarge, yu-etal-2025-table} reason flexibly over serialized tables and documents, but bypass symbolic query execution, leading to high inference cost and poor scalability on large datasets \cite{zhang2025same}. Recent hybrid pipelines \cite{khoja2025weaver, abhyankar2025h, patel2025semantic} combine symbolic querying with semantic matching, but typically operate at a coarse granularity and assume that relevant schema attributes are already explicitly represented. They simplify the problem to attribute identification rather than dynamic structure recovery.

In contrast, we focus on a more fundamental challenge: environments where query-essential attributes are unmaterialized and must be dynamically recovered from unstructured text during query execution.
We argue that this requires moving beyond pipeline-based designs toward a unified execution model that tightly interleaves symbolic database operations with semantic reasoning. This unified approach naturally unlocks fine-grained, data-driven optimizations, allowing the resulting system to generalize effectively across long tables, multi-table schemas, complex queries, textual columns, and implicit attributes. Table~\ref{tab:tqa_comparison} summarizes the full scope of settings we aim to support.

To this end, we propose \System, a unified framework that treats semantic reasoning as a first-class operator in a relational execution model. It translates natural language queries into directed acyclic graphs (DAGs) of atomic relational and LLM-based operators to enable fine-grained reasoning over hybrid schemas. To handle natural language and textual ambiguity when recovering latent schema elements, \System employs a parallel plan diversification strategy. Rather than committing to a single fragile grounding, the planning layer generates and evaluates multiple candidate plans in parallel \seiji{and then consolidates their outputs to select the best answer}.




To execute these heterogeneous plans efficiently, \System incorporates a rule-based query optimization layer and a physical dual-engine architecture. It leverages the operator-centric DAG abstraction and applies classic database optimization principles such as filter push-down, operator reordering, and query-aware reduction to minimize expensive LLM invocations. The dual-engine runtime then dynamically routes these optimized operators between a traditional relational database engine and an LLM module. To ensure throughput over large datasets, the engine implements operator-aware batching and partitions intermediate relations to execute semantic operations in parallel.



We evaluate \System on both structured and semi-structured benchmarks. The results demonstrate that our framework breaks the traditional trade-off. It significantly outperforms existing TQA systems on semi-structured datasets while remaining competitive on fully structured datasets.

Our main contributions are as follows:
\begin{itemize}[leftmargin=*]
  \item We formalize hybrid query processing over semi-structured data, where latent schema elements must be dynamically recovered before symbolic execution can proceed.
  
  \vspace{-5pt}
  \item 
  We design and implement an operator-centric TQA framework that models hybrid queries as relational-semantic DAGs, powered by a physical dual-engine runtime with operator-aware batching for scalable inference.
  
  \vspace{-5pt}
  \item We introduce a parallel plan diversification strategy to robustly recover latent structure under textual ambiguity, and showcase extensibility of the framework via an initial query optimizer.


    \vspace{-5pt}
    \item We conduct extensive experiments showing that \textsc{\System} outperforms existing TQA systems on semi-structured datasets by 14 accuracy points on average and 27 points on the most challenging subset, while maintaining competitive performance on structured benchmarks.
\end{itemize}

\begin{table}[t]
\centering
\renewcommand{\arraystretch}{1.2}
\resizebox{\columnwidth}{!}{
\begin{tabular}{@{} l|cccc @{}}
\toprule
\textbf{Challenge} & \textbf{NL2SQL} & \textbf{Direct-LLM} & \textbf{Hybrid} & \textbf{\System} \\ 
\midrule
Long tables       & \cmark  & \xmark  & \xmark  & \cmark \\ 
Multi-table       & \cmark  & \cmark  & \xmark  & \cmark \\
Complex queries   & \cmark  & \xmark  & \xmark  & \cmark \\
Textual columns   & \xmark  & \cmark  & \cmark  & \cmark \\
Implicit schema   & \xmark  & \cmark  & \xmark  & \cmark \\ 
\bottomrule
\end{tabular}
}
\caption{Comparison of various TQA approaches. \xmark{} denotes settings that are not natively supported, although limited or indirect handling may still be feasible.}
\label{tab:tqa_comparison}
\end{table}

\section{Preliminaries}
\label{sec:background}

\paragraph{Data Model.} Let $\mathcal{S}=\{T_1, \dots, T_n\}$ be a database schema over an instance $\mathcal{D}$. The global attribute set $\mathcal{A} = \bigcup_i \mathcal{A}_i$ is partitioned into two disjoint subsets: structured attributes $\mathcal{A}_{\mathrm{s}}$ and unstructured text attributes $\mathcal{A}_{\mathrm{u}}$, such that $\mathcal{A} = \mathcal{A}_{\mathrm{s}} \cup \mathcal{A}_{\mathrm{u}}$. Structured attributes contain atomic values queryable via traditional relational operators $\Omega_{\mathrm{R}}$ (e.g., projection, selection, join, aggregate).
Unstructured attributes contain free-form text fields that implicitly encode latent entities, attributes, or predicates, requiring a set of semantic operators $\Omega_{\mathrm{S}}$ (e.g., extraction, matching, joins).

\paragraph{Table Question Answering.}
Given a structured schema ($\mathcal{A} = \mathcal{A}_{\mathrm{s}}$) and a natural language query $Q$, the objective is to derive a symbolic execution plan composed exclusively of relational operators. Executing this plan over $\mathcal{D}$ yields the answer $Y$.

\paragraph{Semi-Structured Table Question Answering.}
Given a hybrid schema ($\mathcal{A} = \mathcal{A}_{\mathrm{s}} \cup \mathcal{A}_{\mathrm{u}}$) and a query $Q$, the objective is to dynamically recover the latent structure embedded within $\mathcal{A}_{\mathrm{u}}$ and ground it alongside explicit elements in $\mathcal{A}_{\mathrm{s}}$. A hybrid plan of relational and semantic operations is required to jointly reason over $\mathcal{A}_{\mathrm{s}}$ and $\mathcal{A}_{\mathrm{u}}$.


\subsection{Design Considerations}
\label{ssec:design_principles}


To scale across long tables, multi-table schemas, complex queries, textual columns, and implicit attributes (Table~\ref{tab:tqa_comparison}), a hybrid query processing system should adhere to three core principles:

\begin{itemize}[leftmargin=*]
  \item \textbf{Operator-Centric Decomposition.} Decompose queries into fine-grained plan with sub-tasks that can be directed to a symbolic or semantic engine. Such decomposition is essential to handle complex queries and multi-table operations over implicit text attributes.
   
\item \textbf{Plan Diversification.} Mitigate natural language and text column ambiguity by generating and exploring alternative logical plans. This ensures robust structural grounding before executing downstream database operations.

  
  \item \textbf{Cost Optimization.} 
  Apply principles from database query optimization to reduce expensive LLM inferences and allow the framework to scale to long tables efficiently.
  
\end{itemize}

\section{Proposed System: \System}
\label{sec:System}

\begin{figure*}[t]
  \centering
  \includegraphics[width=0.95\textwidth]{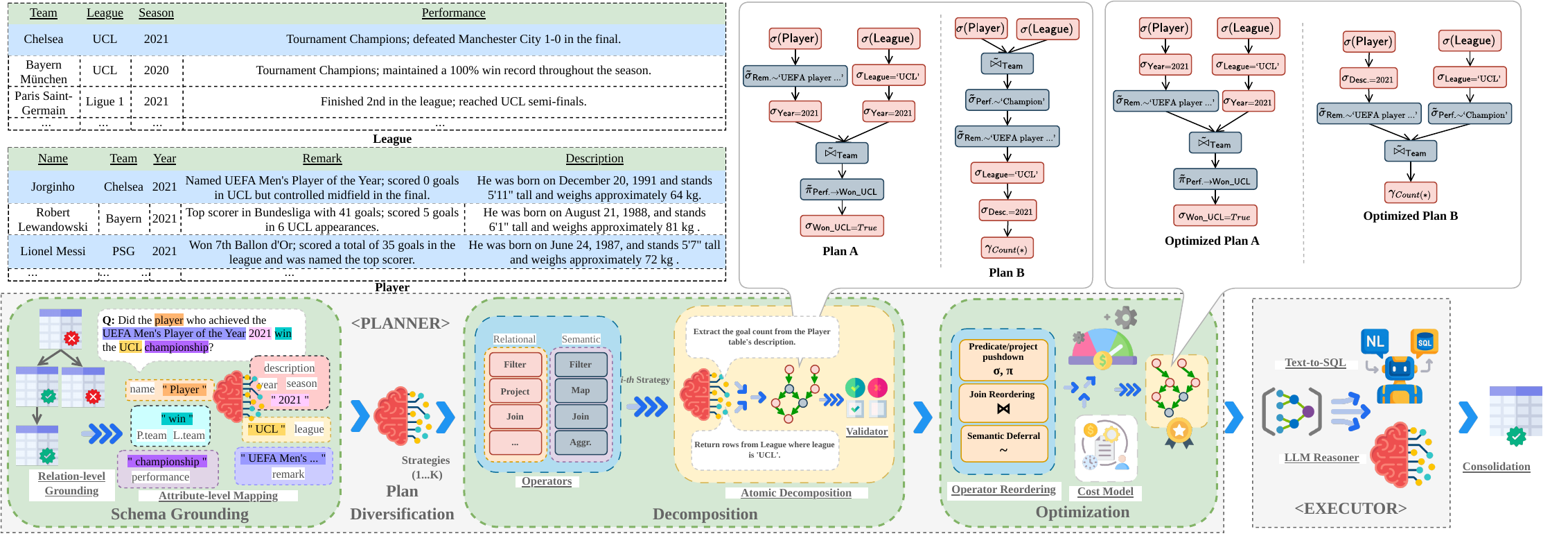}
    \caption{\System architecture. Given the \textsf{UEFA Soccer} database and the query \textit{``Did the player who achieved the UEFA Men's Player of the Year 2021 win the UCL championship?''}, \System constructs a data preview (blue) to ground query expressions to schema elements, then generates and optimizes candidate logical plans to resolve ambiguity over where the UEFA Men's Player of the Year information is stored, e.g., in ``Remark'' or ``Description''.}
  \label{fig:planning}
\end{figure*}

To operationalize these principles, we introduce \System, which cleanly decouples query planning 
from physical execution. We illustrate the overall architecture of \System in Figure~\ref{fig:planning}, which consists of a logical planning layer with plan diversification, and a physical dual-engine runtime. 


\subsection{Pre-processing}

Given a natural language query $Q$, 
\System first executes a lightweight, query-aware pre-processing to prune the database context and build a compact data profile for the planner. This stage implements two primitives. First, an LLM isolates query-relevant tables, explicit attributes $\mathcal{A}_{\mathrm{s}}$, and unstructured text fields $\mathcal{A}_{\mathrm{u}}$, while eliminating orthogonal schema elements to minimize the plan search space. Next, it constructs a representative data snapshot for the planner by combining semantically relevant rows (via vector search over text fields) with randomly sampled tuples. This provides the planner grounded context regarding the underlying data distribution. We provide implementation details of pre-processing in Appendix~\ref{app:sec:preprocessing}. The output of this stage consists of a pruned database context and a query-aware data preview, which  together serve as input for the planning stage.


\subsection{Logical Planning \& Plan Diversification}
\label{ssec:planning}
The logical planning phase consists of four components: \textit{schema grounding}, \textit{atomic decomposition}, \textit{plan diversification}, and \textit{plan optimization}. 

\paragraph{Schema Grounding.}Building directly on the preprocessing layer, {schema grounding} utilizes the dual data preview to map the entities and predicates expressed in $Q$ to concrete elements of the refined database schema $\mathcal{S}$ (e.g., mapping the phrase ``customer feedback'' to a specific text column named \texttt{usr\_cmnt}). This process resolves linguistic variations and ensures all query-relevant tables and attributes are bound. 



\paragraph{Operator-Centric Decomposition.}

Once the relevant schema elements are identified, the planner decomposes $Q$ into a directed acyclic graph (DAG) representing a logical query plan, denoted as $\mathcal{G} = (\mathcal{V}, \mathcal{E})$. Each node $v \in \mathcal{V}$ corresponds to an atomic operator drawn from the unified operator space $\Omega = \Omega_{\mathrm{R}} \cup \Omega_{\mathrm{S}}$, while directed edges $e \in \mathcal{E}$ define strict data-flow dependencies. To ensure fine-grained execution and deterministic optimization, we require every atomic node $v$ to satisfy two structural constraints: (1) it contains at most one logical condition, and (2) the condition involves a single column or variable.

By modeling queries as explicit operator DAGs, \System extends classical query optimization to semantic spaces. Unlike prior hybrid pipelines \cite{khoja2025weaver, abhyankar2025h} where semantic operators only process pre-defined schema attributes, nodes in $\Omega_{\mathrm{S}}$ are dynamically tasked with recovering latent relational structures from text fields $\mathcal{A}_{\mathrm{u}}$. They can convert unstructured text into structured relations to unblock downstream symbolic operations $v \in \Omega_{\mathrm{R}}$. We provide the details of relational $\Omega_{\mathrm{R}}$ and semantic $\Omega_{\mathrm{S}}$ operators in Table~\ref{tab:operators} in Appendix \ref{app:sec:operators}.



\paragraph{Plan Diversification.}

To mitigate structural failures arising from natural language ambiguity or unmaterialized schemas, \System implements a parallel \textit{plan diversification} strategy. Rather than generating a single, error-prone schema grounding, the planner generates a set of $K$ distinct candidate execution plans. These candidate plans systematically explore alternative structural interpretations across four primary dimensions:
  \textbf{1) Schema Mapping Diversity:} We map an ambiguous query term such as ``return player'' to multiple candidate attributes such as \texttt{player\_id} and \texttt{player\_name}.
  \textbf{2) Risk-Profile Variations:} We generate \textit{strict} plans that prioritize deterministic relational operators, as well as \textit{fuzzy} plans that use semantic operators to capture subjective predicates.
  \textbf{3) Operator Substitution:} We explore logically equivalent transformations, such as using a semantic \texttt{FILTER} instead of a semantic \texttt{MAP} followed by relational filtering.
  \textbf{4) Semantic Intent Modeling:} We interpret vague expressions such as ``top player'' using alternative semantic criteria, such as \texttt{goals\_scored} or \texttt{awards\_won}.


\paragraph{Plan Optimization.}

The execution plans generated by the planner may not be cost-efficient, particularly when semantic operators are involved. Because query optimization over hybrid operator spaces is NP-hard~\cite{chatterji2002complexity}, we employ a LLM-based heuristic optimizer that accounts for the asymmetry between inexpensive relational operations and expensive semantic reasoning. We enforce four optimization invariants over the operator DAGs:
\textbf{1) Selection Pushing}: Relational selection operators ($\sigma$) are pushed down past semantic operators toward base relations to prune irrelevant tuples early. \textbf{2) Projection Pruning}: Unnecessary attributes are dropped immediately after their last dependent node to reduce the width and memory footprint of intermediate relation. \textbf{3) Join Reordering}: Highly selective relational joins are prioritized to limit intermediate result size.  \textbf{4) Adaptive Semantic Deferral}: Semantic operators are systematically deferred until relational pruning is complete  unless a downstream relational operator is expected to expand the result set, in which case the semantic step is moved earlier to serve as a pre-filter. We also provide a deterministic implementation of this optimization strategy in Appendix~\ref{app:sec:plan_optimization_details}.

\subsection{Physical Execution \& Plan Consolidation}
\label{ssec:execution}

The physical execution engine takes the $K$ optimized candidate plans and executes them concurrently over the database instance $\mathcal{D}$ to produce $K$ candidate answers, which are subsequently consolidated into a single final output. To balance structural scaling with semantic expressiveness, the runtime introduces a dual-engine architecture. Relational operators ($\Omega_{\mathrm{R}}$) are dynamically routed to a native SQL database engine that executes compiled query statements directly on the host database. Conversely, semantic operators ($\Omega_{\mathrm{S}}$) are dispatched to an LLM-based reasoning module. To prevent the context-window explosions or excessive row-wise API latency that a naive data-passing approach would incur, the engine employs a block-based batching strategy.

\paragraph{Operator-driven Batching.}
The execution engine treats semantic operators as chunked primitives to enable substantial intra-operator parallelism. Intermediate relations are partitioned into manageable blocks of size $\beta$ and executed concurrently via three asynchronous batching strategies:
 \textbf{1) Semantic Map} \& \textbf{Filter}: 
    {These operators evaluate tuples independently. Input relations are partitioned into chunks processed asynchronously. To reduce output tokens, the LLM returns only derived values (for {Map}) or row indices satisfying the predicate (for {Filter}), which are then applied to the original relation.}    
  \textbf{2) Semantic Join}: 
    {To avoid loading massive Cartesian products into a single context window, the engine executes an LLM-driven block nested-loop join. Input relations are streamed in blocks of size $\beta$, and each block pair is evaluated independently in parallel.}
  \textbf{3) Semantic Aggregate}: 
    {For operators requiring global context, the engine employs a recursive map-reduce strategy. Relations exceeding $\beta$ are partitioned into parallel chunks, partial local aggregations are computed asynchronously, and intermediate metrics are recursively reduced to a final scalar value.}
The detailed algorithm for operator-driven batching is described in Appendix~\ref{app:sec:batch_execution_details}.

\paragraph{Plan Consolidation.}
The final stage of the execution step is to select the best answer among the $K$ candidates produced by executing the diversified plans.
\System supports two consolidation strategies: 
	\textbf{1) LLM-as-a-Judge Consensus}~\cite{gu2024survey}. An independent LLM evaluates candidate outputs by comparing them against the original query and representative samples of each plan's results. Using few-shot examples to calibrate the evaluation, the judge selects the output that best aligns with the user's intent.
	\textbf{2) Semantic Majority Voting}~\cite{chen2024more}. An ensemble-based strategy that treats each materialized relation as a vote and selects the result that appears most frequently across independent plans.

\section{Experiments}
\label{sec:experiments}

We aim to answer four core research questions:

\begin{itemize}[leftmargin=*]
\item \textbf{RQ1 (Effectiveness):} Does \System outperform state-of-the-art baselines on hybrid schemas containing implicit text attributes?
\item \textbf{RQ2 (Generalizability):} Does the performance generalize across  datasets and base LLMs? 
\item \textbf{RQ3 (Efficiency):} How does \System compare to baselines in terms of token cost?  
\item \textbf{RQ4 (Ablation Study):} What is the impact of individual design components?
\end{itemize}

\paragraph{Datasets.}
Our primary benchmark is \textsc{RepairTQA} \cite{zhang2025same}, which provides paired fully structured and semi-structured variants of the same underlying tables. In the semi-structured setting, query attributes embedded in free-form text fields.
It also includes subsets with simple lookup, compositional queries, and multi-table operations across both short and long table settings.
To evaluate generalization, we additionally evaluate on semi-structured datasets including \textsc{FetaQA} \cite{nan2022fetaqa}, \textsc{HybridQA} \cite{chen2020hybridqa}, and \textsc{TAT-QA} \cite{zhu2021tatqa} datasets. Detailed dataset statistics are provided in Appendix~\ref{app:sec:dataset_details}. We also evaluate standard fully structured benchmarks including \textsc{WikiSQL} \cite{zhong2017seq2sql} and \textsc{WikiTableQuestions} \cite{pasupat2015compositional}, with results reported in Appendix~\ref{app:sec:conventional_tqa}. 

\paragraph{Baselines.}
We evaluate \System against three categories of baselines: direct-LLM, Text-to-SQL, and hybrid approaches.
\textbf{\textsc{Direct-LLM}} 
performs direct prompting without explicit planning or symbolic execution. \textbf{\textsc{NL2SQL}} generates SQL queries for direct database execution. \textbf{\textsc{Plan-of-SQLs}}~\cite{nguyen2025pos} decomposes a query into natural language steps, translates each step to SQL, and chains intermediate results.  \textbf{\textsc{H-Star}}~\cite{abhyankar2025h} routes operations between SQL and LLMs based on task types, but assumes explicit schema attributes and does not support multi-table reasoning. \textbf{\textsc{Weaver}}~\cite{khoja2025weaver} interleaves SQL execution with LLM reasoning, but assumes explicit schemas and does not target multi-table or semi-structured settings.

\begin{figure*}[t] 
\centering
    \includegraphics[width=\textwidth]{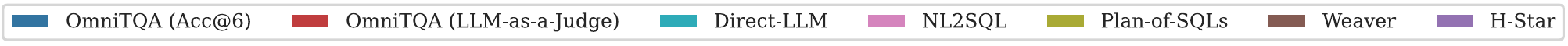}\\
    \begin{minipage}[t]{\linewidth}
        \centering
        \includegraphics[width=0.95\textwidth]{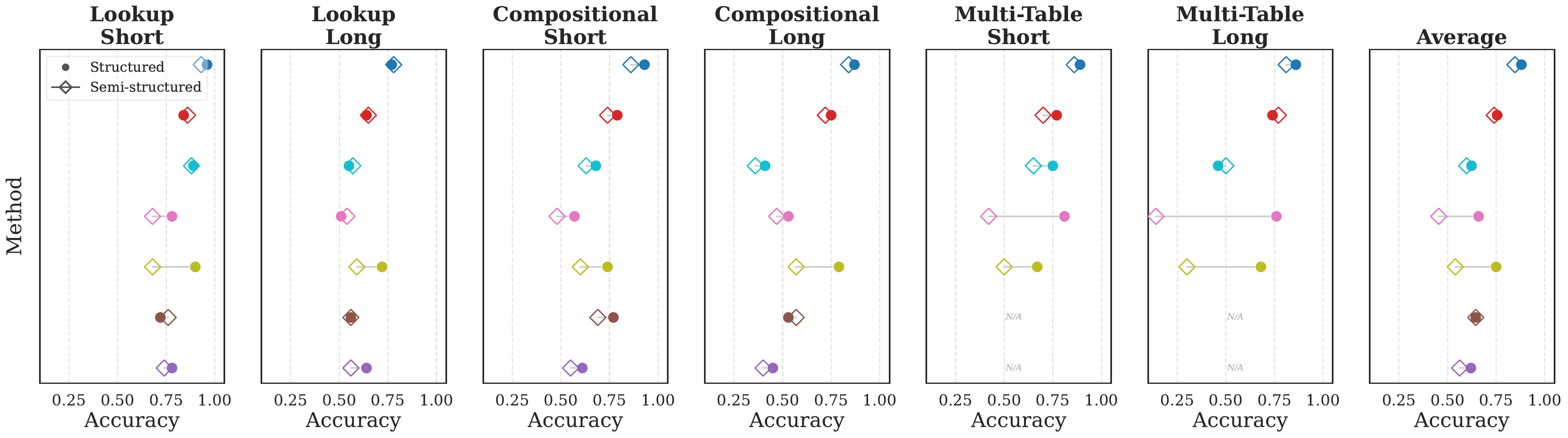}
        \vspace{-.2em}
        \caption{Effectiveness comparison of \System vs. baselines on structured vs. semi-structured representations of \textsc{RePairTQA} evaluated with \textsc{Gemini-3-Flash}. \textbf{Filled markers} denote the structured setting, and \textbf{unfilled markers} denote the semi-structured setting.}
        \label{fig:acc_gemini}
        \vspace{.3em}
    \end{minipage}
    \hfill
    \begin{minipage}[t]{0.21\linewidth}
        \centering
        \includegraphics[width=\textwidth]{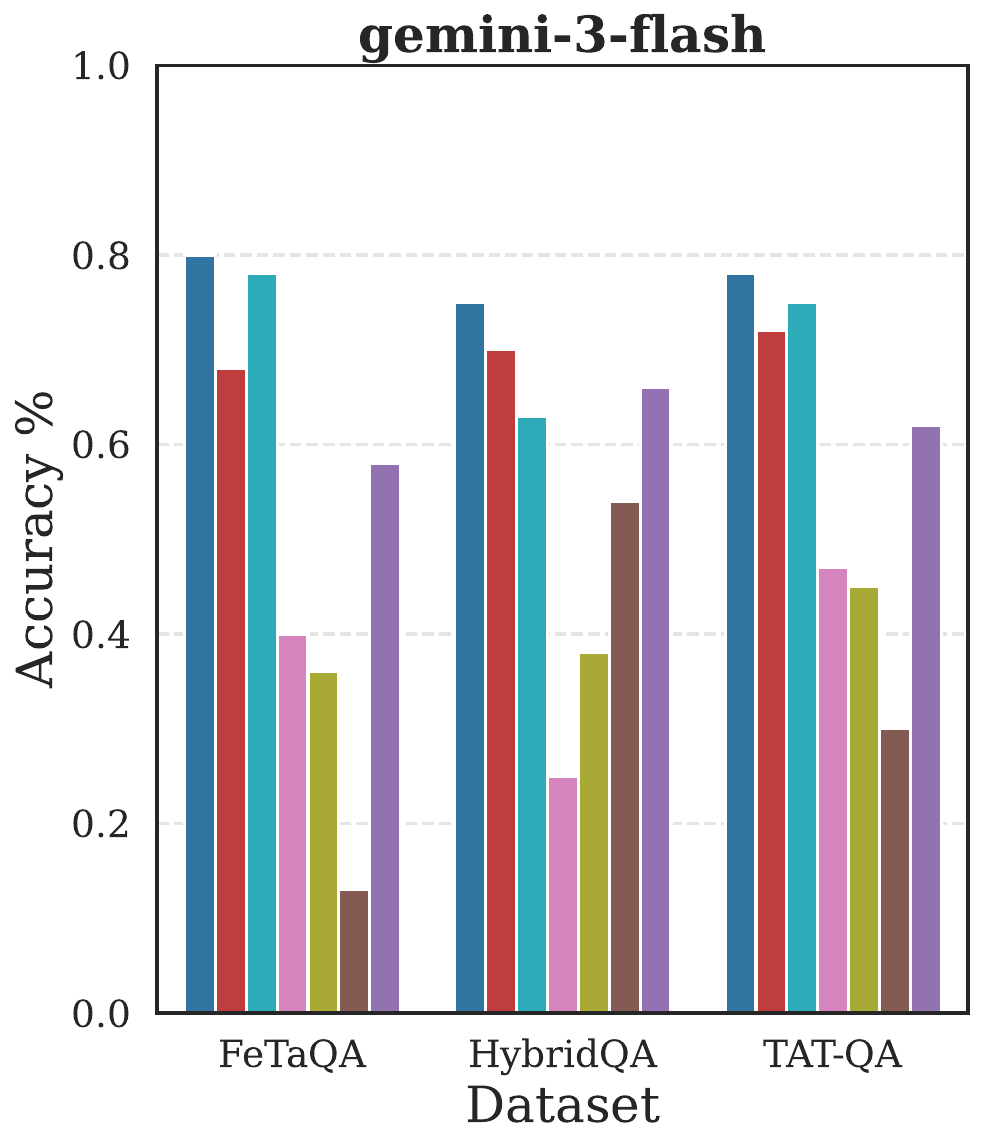}
        \vspace{-1.em}
        \caption{Effectiveness on extended benchmarks.}
        \label{fig:acc_gemini_extended}
    \end{minipage}
    \hfill
    \begin{minipage}[t]{0.28\linewidth}
        \centering
        \includegraphics[width=\textwidth]{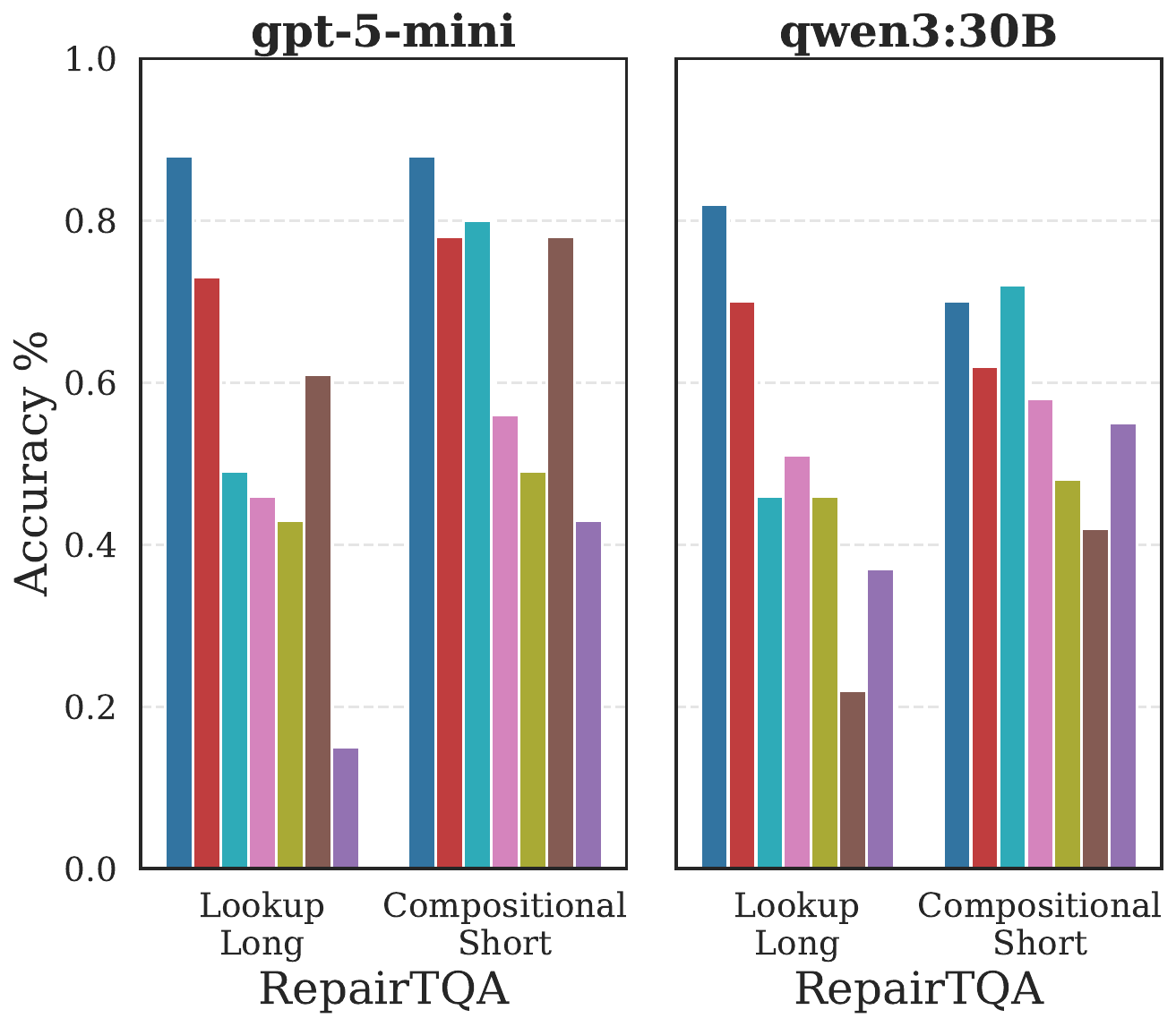}
        \vspace{-1.em}
        \caption{Effectiveness of of \System vs. baselines using different base LLMs.}
        \label{fig:accuracy_vaying_llm}
    \end{minipage}
    \hfill
    \begin{minipage}[t]{0.46\linewidth}
        \centering
        \includegraphics[width=\textwidth]{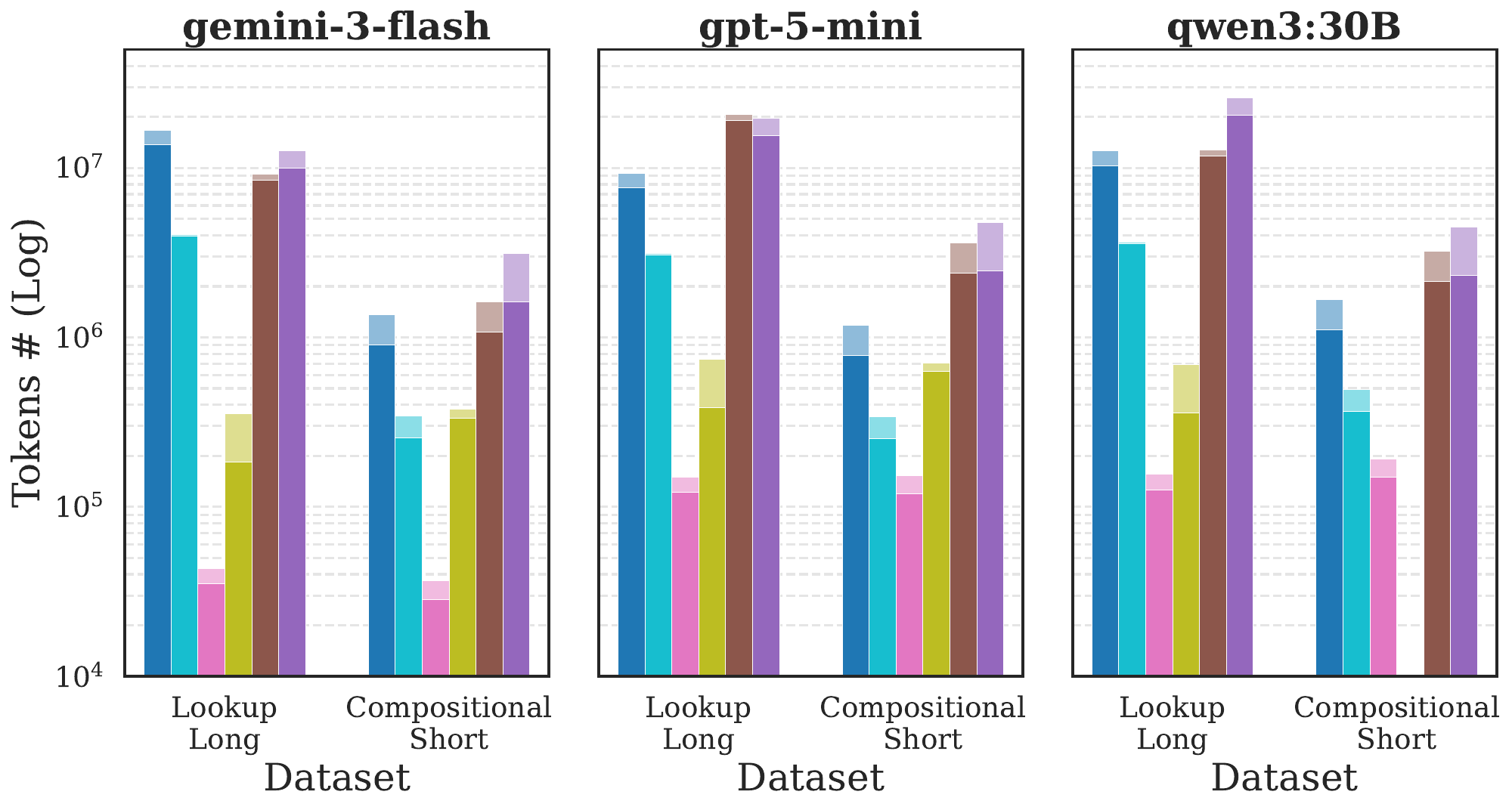}
        \vspace{-1.em}
        \caption{Cost comparison of \System vs. baselines. Darker and lighter shades denote \textit{input} and \textit {output} tokens, respectively.}
        \label{fig:cost}
    \end{minipage}
\end{figure*}

\paragraph{Evaluation. }
Following \textsc{RepairTQA}~\cite{zhang2025same}, we use an LLM-based judge instead of exact string matching to evaluate outputs. This accounts for variations in ordering, formatting, and minor textual differences while preserving semantic equivalence for free-form answers.
We evaluate systems along two dimensions: \textit{effectiveness} (answer accuracy) and \textit{efficiency} (token usage). 

\paragraph{Settings.}
We evaluate across multiple base LLMs, including {Gemini-3-Flash-Preview}, {GPT-5-Mini}, and a locally deployed {Qwen3:30B} via {Ollama}. Vector embeddings are generated using the \texttt{all-MiniLM-L6-v2} encoder and cosine similarity is used for retrieval during pre-processing. We provide additional model details in Appendix~\ref{app:sec:models}. 
By default, we set the plan diversification factor to $K=6$ and the execution block size to $\beta=100$; sensitivity analyses for $K \in [1,6]$ and $\beta \in [10,1000]$ are provided in Appendix~\ref{app:sec:hyperparameter_analysis}. Unless stated otherwise, we use {Gemini-3-Flash-Preview} for all experiments. For fair comparison, all methods use the same underlying LLMs.
\seiji{We report the results of two OmniTQA's plan consolidation settings}: (1) \textsc{Acc@6}, which considers an example correct if any of the $K$ plans produces the correct answer\footnote{\textsc{Acc@6} reflects the upper-bound performance achievable under ideal plan consolidation.}; and (2) {LLM-as-a-Judge}, where an independent LLM selects the best answer among the $K$ candidates.\footnote{We discuss other consolidation variants in Appendix~\ref{app:sec:alt_consolidation}} 
Additional implementation details and prompts are provided in Appendix~\ref{app:sec:preprocessing} and Appendix~\ref{app:sec:prompts}, respectively.

\subsection{RQ1: Effectiveness}
\label{ssec:effectiveness}


We first evaluate the methods on structured and semi-structured subsets in \textsc{RepairTQA} to measure their effectiveness in recovering latent structure. 


\paragraph{\System achieves state-of-the-art accuracy across semi-structured settings.} As shown in Figure~\ref{fig:acc_gemini}, \System consistently outperforms all baselines. Using LLM-as-a-Judge, it outperforms the strongest baseline (Direct-LLM) by {14 points} on average in the semi-structured setting (74\% vs.\ 60\% in the rightmost subplot). The largest gain appears on the multi-table long subset, where \System achieves {77\%} accuracy compared to {50\%} for the next-best baseline (Direct-LLM). This setting is particularly challenging because it requires both accurate schema grounding and complex join reasoning over implicit attributes. These gains suggest that plan diversification helps latent structure under ambiguity. While LLM-as-a-Judge does not reach the \textsc{Acc@6} upper bound, it effectively selects strong candidates from the diversified plan set and consistently outperforms all baselines.



\paragraph{Direct-LLM struggles on long tables, while Text-to-SQL degrades sharply on semi-structured data.} Baseline architectures fail under specific table topologies. Direct-LLM remains competitive on when serialized tables fit comfortably within the model context window. Conversely, Text-to-SQL baselines excel on explicit schemas but struggle when query-relevant attributes are embedded in free-form text. This degradation becomes more severe on multi-table subsets that require both schema grounding and join reasoning.


\paragraph{Hybrid baselines shows moderate robustness but remain limited on complex settings.} Hybrid baselines outperform Text-to-SQL systems on semi-structured data, but still struggle on long-table and multi-table subsets. This suggests that coarse-grained integration between symbolic execution and semantic reasoning is insufficient for robust latent structure recovery at scale.


\paragraph{\System significantly narrows the performance gap between structured and semi-structured settings.}
All methods experience some degradation when moving from structured to semi-structured tables. However, \System exhibits the smallest performance drop, indicating stronger robustness to implicit schema representations. This suggests that its unified planning and execution framework more effectively adapts to latent structure embedded within unstructured text.

\providecommand{\vhead}[1]{\rotatebox[origin=c]{90}{\textbf{#1}}}

\begin{table}[t]

\centering
\resizebox{\columnwidth}{!}{
\begin{tabular}{l|c|ccccc}
	\toprule
	\textbf{Dataset} & \vhead{\System} & \vhead{w/ Naive $\Preview$} & \vhead{w/ QDMR} & \vhead{w/o Opt. $\mathcal{G}$} & \vhead{w/o Div. $\mathcal{G}$} & \vhead{w/o Pru. $\Schema$} \\ \midrule
Lookup Long & \textbf{0.65} & \cellcolor{moddrop}0.59 & \cellcolor{sigdrop}0.43 & 0.65 & \cellcolor{moddrop}0.53 & 0.65 \\ \hline
Composional Short & \textbf{0.74} & 0.74 & \cellcolor{sigdrop}0.46 & 0.74 & \cellcolor{moddrop}0.68 & \cellcolor{slidrop}0.73 \\ \hline
Multi-Table Short & \textbf{0.70} & \cellcolor{moddrop}0.64 & \cellcolor{sigdrop}0.33 & 0.70 & \cellcolor{moddrop}0.59 & \cellcolor{slidrop}0.68 \\
\bottomrule
\end{tabular}
}
\caption{Ablation study: accuracy (LLM-as-a-Judge).}
\label{tab:ablation_llm_judge}
\end{table}

\begin{table}[t]
\centering
\resizebox{.85\columnwidth}{!}{
\begin{tabular}{l|c|c}
	\toprule
	\textbf{Dataset} & \textbf{\System} & \textbf{w/o Opt. $\mathcal{G}$} \\ \midrule
Lookup Long & \textbf{16.7M} & \cellcolor{moddrop}20.4M \\ \hline
Composional Short & \textbf{1.3M} & \cellcolor{moddrop}1.6M \\ \hline
Multi-Table Short & \textbf{0.6M} & \cellcolor{slidrop}0.7M \\
\bottomrule
\end{tabular}
}
\caption{Ablation study: cost (LLM-as-a-Judge).}
\label{tab:ablation_cost}
\end{table}

\subsection{RQ2: Generalizability}
\paragraph{\System generalizes well across semi-structured benchmarks.}
Figure \ref{fig:acc_gemini_extended} shows the accuracy of \System and baselines on three additional semi-structured datasets. Because these benchmarks contain relatively small tables, Direct-LLM performs reasonably well. However, Text-to-SQL methods fail due to the implicit schema representations.
Overall, \System consistently achieve strong performance across all datasets. 
\vspace{-5pt}

\paragraph{\System is robust across different base LLMs, particularly on long-table subsets.}
Figure \ref{fig:accuracy_vaying_llm} compares method performances with  GPT-5-Mini and Qwen3:30B. \System shows similar trends as Gemini-3-Flash-Preview, with strong gains on long-table subsets, indicating robustness to the choice of base LLM. In contrast, hybrid baselines exhibit larger performance variation across LLMs, suggesting higher sensitivity to model capabilities and prompting strategies. \System's modular architecture and parallel plan diversification seem to provide a more stable performance across different LLMs.

\subsection{RQ3: Efficiency}
\label{ssec:efficiency}
We evaluate computational cost in terms of token usage. Figure~\ref{fig:cost} reports input and output tokens (stacked bars) with darker segments indicating input tokens and lighter segments output tokens. We compare configurations using one versus six candidate plans. LLM-as-a-Judge has similar cost to the six-plan setting and is omitted for clarity. All results are averaged over the full test set.


\paragraph{\System achieves achieves superior accuracy and robustness at competitive cost.}
A single-plan configuration of \System has cost comparable to Direct-LLM, while delivering substantially higher accuracy and robustness. 
In contrast, Text-to-SQL methods, especially NL2SQL, are the most cost-efficient due to minimal LLM involvement. Finally, we observed that across all settings, input tokens dominate the total cost.
\vspace{-5pt}

\subsection{RQ4: Ablation Study}
\label{ssec:ablation}

We conduct an ablation study of \System's key components. While we focus here on accuracy via LLM-as-a-Judge, identical relative trends hold \textsc{Acc@6} (see Appendix Table~\ref{tab:ablation_acc6_appendix}).


\vspace{-5pt}  
\paragraph{Data preview is essential for long tables.} 
Replacing data preview with a uniformly sampled subset consistently degrades performance, particularly on long tables (Table~\ref{tab:ablation_llm_judge}). The effect is smaller on short tables, where relevant information is more likely to appear in random samples.


\vspace{-5pt}
\paragraph{Operators designed for semi-structured settings are crucial for accurate decomposition.}
Replacing our operator model with a QDMR-based decomposition~\cite{wolfson-etal-2020-break} results in significant accuracy degradation across all settings, reaching as high as 49\%. This highlights that the design of operators for semi-structured table reasoning is a non-trivial task. 

\paragraph{Plan optimization improves efficiency without affecting accuracy.}
Removing optimization increases execution cost by up to 20\% ( Table \ref{tab:ablation_llm_judge}), while leaving accuracy unchanged. This suggests that our optimization strategy successfully minimize inference overhead without modifying the plan's underlying semantics.

\paragraph{Plan diversification is necessary for robust performance.} 
Replacing our diversification strategy with naive sampling of $K$ independent plans leads to consistent accuracy degradation. This suggests that stochastic LLM variation is insufficient for handling schema ambiguity or structural recovery. 

\paragraph{Schema pruning provides consistent but modest gains.}
Removing schema pruning leads to a modest decrease in accuracy, especially on complex queries. 
\seiji{Table \ref{tab:ablation_cost} shows that pruning also reduces token cost by up to 18\%. 
This suggests that pruning effectively reduces the search space for the planner, leading to efficient and accurate plan generation.}

\begin{figure}[t]
\centering
  \includegraphics[width=.9\linewidth]{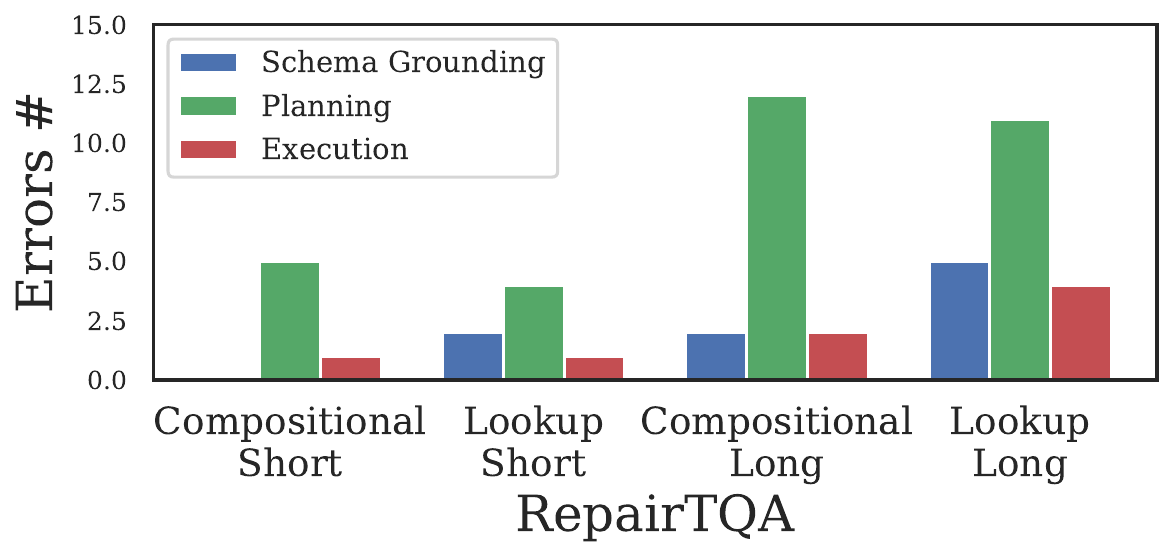}
  \caption{Error categorization in \textsc{RepairTQA}.}
  \label{fig:error_analysis}
\end{figure}

\subsection{Error Analysis}
\label{ssec:error_analysis}
We conduct a qualitative analysis of failure cases to identify sources of error. While some errors in \textsc{RepairTQA} stem from data inconsistencies or evaluation issues (detailed in Appendix~\ref{app:sec:failure_cases}), we focus on errors attributable to the system itself. We categorize failures into three stages of \System: \textit{schema grounding}, \textit{planning}, and \textit{execution}, based on manual inspection of errors in single-table subsets of \textsc{RepairTQA}.


Figure~\ref{fig:error_analysis} shows that the majority of errors are due to planning, which includes decomposition and plan generation. 
Common issues include incomplete projections (e.g., missing requested attributes such as middle name when `full name' is required). Because the planner relies on a compact data snapshot, it occasionally generates filters using misaligned semantic literals (e.g., substituting {status}=`extinct' with `inactive' or `disappeared'). 
This also leads to incorrect operator choices. For example,  it incorrectly maps `Northern California' to a {region} column value that is not present in the data, or treats `most verbose complaint' as word count rather than {service duration}.
We provide additional examples in  Appendix~\ref{app:sec:error_analysis_details}.

\section{Related Work}
\label{sec:related_works}
\vspace{-5pt}
\seiji{
\noindent \textbf{Text-to-SQL and Table QA.}
Text-to-SQL methods execute efficiently over structured databases \cite{luo2025survey,hong2025next,liu2024survey,deng2025reforce}, but they assume explicit schemas and struggle when query-relevant information is embedded in text fields. LLM-based TQA methods instead serialize tables and text for direct reasoning \cite{zhang2025survey,yu-etal-2025-table,wu2025tablebench,wu2025mmqa}, offering greater semantic flexibility but sacrificing relational efficiency and scalability on long or multi-table settings.

\noindent \textbf{Semantic Operator Systems.}
Recent database systems integrate LLMs as fine-grained semantic operators \cite{patel2025semantic,liu2025palimpzest,glenn2024blendsql,su2026large}, while hybrid pipelines route sub-tasks between relational engines and LLMs \cite{khoja2025weaver,abhyankar2025h}. 
However, both assume that schema attributes and relational linkages are explicit, making them unsuitable when the attributes or join conditions needed for execution remain latent in text.
}

In contrast, \textsc{\System} targets environments where query-relevant schemas are unmaterialized and must be dynamically reconstructed during query execution~\cite{chen2020hybridqa, zhang2025same}.  Rather than utilizing semantic operators merely as downstream data-enrichment functions or isolated pipeline steps, \textsc{\System} treats them as primitive nodes within a DAG. This operator-centric design enables plan diversification for robustly resolving schema ambiguity and plan optimizations for scalable execution.

\section{Conclusion}
\label{sec:conclusion}
\vspace{-5pt}
We presented \textsc{\System}, a unified framework for hybrid query processing over semi-structured data where query-relevant structure must be dynamically recovered during execution. \System translates queries into directed acyclic graphs (DAGs) that combine relational and semantic operators. This enables cost-aware planning, dual-engine execution, and plan diversification to robustly recover latent schema elements while maintaining scalability. Our experiments show that \System consistently sets a new accuracy standard on semi-structured benchmarks while maintaining low token costs and high accuracy on structured settings.



\section{Limitations}
\label{sec:limitations}
Despite these advances, challenges remain in planning quality, cost, and latency. In particular, generating diverse high-quality plans remains a bottleneck. Since optimization is inherently limited by the quality of the initial plan, incorporating human-in-the-loop to audit and prune plans could significantly improve reliability and cost-efficiency. Furthermore, semantic operators continue to incur significant overhead. A tiered execution approach involving a high-recall, efficient hash-based \textsf{FILTER} or \textsf{JOIN} to prune the search space prior to the more expensive LLM-based execution can be helpful. This can be further augmented by model cascading strategies, which utilize a hierarchy of models to route simpler logic to smaller, cost-effective models while reserving larger LLMs for complex reasoning tasks.

\section{Ethical Considerations}
\label{sec:ethical_considerations}
We made use of AI tools such as ChatGPT and Copilot to support coding and refining this paper, but all content was carefully reviewed and edited by us to ensure it adheres to our standards and aligns with our research objectives.
All external artifacts, datasets, and models utilized in this study were used in strict accordance with their original intended purposes and respective licensing agreements, adhering to established ethical guidelines for research.

\bibliography{ref}

\appendix

\section{Further Deep Dive into \System's Components}
\label{app:sec:omnitqa_details}

\begin{table*}[t]
\centering
\resizebox{\linewidth}{!}{
\begin{tabular}{lll}
\toprule
\textbf{Attribute Category}  & \textbf{Relational Type Binding}  & \textbf{Extracted Profile Summaries} \\ \midrule
\textbf{Numeric}   & $\{\textit{int, float}\}$ & Min, Max, Average, Variance \\ 
\textbf{Categorical} & $\{\textit{string, enum}\}$ & Cardinality, Top-$k$ frequent values \\ 
\textbf{Textual}   & $\{\textit{varchar, text}\}$ & Min/Max length, unique count, sample snippets, semantic summaries \\ 
\textbf{Temporal}  & $\{\textit{date, timestamp}\}$ & Range [$t_{\textit{start}}, t_{\textit{end}}$], Granularity (e.g., year, day) \\ 
\textbf{Relational} & $\{\textit{PK, FK}\}$ & Referential constraints, join key mappings, table dependencies \\ 
\bottomrule
\end{tabular}
}
\caption{Data Profiling for Schema Augmentation}
\label{tab:schema_augmentation}
\end{table*} 

\subsection{PREPROCESSING}
\label{app:sec:preprocessing}
\System preprocessing is divided into two phases. A query-agnostic phase first normalizes the database instance and enriches the schema before any question is observed. A query-aware phase then constructs a question-specific view of the data that supports downstream planning and execution.

\paragraph{Query-agnostic Preprocessing.}
The query-agnostic phase is executed once per database instance to improve structural reliability and reduce irrelevant context before planning begins. It consists of three operations. First, \textbf{table cleaning and normalization} resolves structural inconsistencies such as duplicate attributes, malformed columns, and noisy fields so that downstream SQL execution remains stable. Second, \textbf{heuristic schema pruning} removes low-value attributes that are unlikely to help question answering, including columns with more than 95\% NULL values, near-constant columns, and clearly non-informative content such as password hashes, Base64 strings, or code fragments. Third, \textbf{schema augmentation} attaches lightweight profiles to the remaining attributes, including inferred types, statistical summaries, and relational constraints such as primary and foreign keys. These profiles provide the planner with stronger relational grounding; Table~\ref{tab:schema_augmentation} summarizes the extracted metadata.

\paragraph{Query-aware Preprocessing.}
The query-agnostic phase improves schema quality, but the planner still needs question-conditioned evidence to align natural language predicates with concrete attributes and values, especially in semi-structured tables where relevant signals may remain implicit in text. To address this, \System constructs a query-aware view tailored to the input question.

This phase has three components. First, \textbf{semantic schema pruning} performs a high-recall LLM pass that keeps attributes likely to be relevant to the query while reducing token overhead. Second, \textbf{schema grounding} maps query entities and predicates to candidate tables and attributes using the pruned schema, attribute profiles, and retrieved examples. Third, a \textbf{query-aware data preview} combines \textit{semantic search} and \textit{random sampling}: semantic search retrieves the top-$k_1$ rows most relevant to the predicates in a question, while random sampling contributes $k_2$ tuples that capture the general format and diversity of the relation. Together, these steps produce a refined database instance and a preview that ground both plan generation and execution.


\subsection{PLANNING: Hybrid Operator Model}
\label{app:sec:operators}
We define a comprehensive set of operators that form the building blocks of \System's execution engine. These operators are categorized into two classes: Relational Operators, which perform traditional database operations, and Semantic Operators, which leverage LLM capabilities for tasks that require understanding and reasoning over unstructured data. Table \ref{tab:operators} provides a detailed overview of each operator, including its function, instruction template, and description.

\begin{table*}[ht]
\scriptsize
\centering
\renewcommand{\arraystretch}{1.4}
\begin{tabular}{l|l|l|p{4cm}|p{6cm}}
\toprule
\textbf{Class} & \textbf{Operator} & \textbf{Function} & \textbf{Instruction Template} & \textbf{Description} \\ \midrule
\multirow{8}{*}{\rotatebox[origin=c]{90}{\textbf{Relational}}} 
 & \textsf{SCAN}      & $\sigma(\Instance)$ & \textit{``Return rows from $\Instance$.''} & Retrieve base relation instance $\Instance_i \in \Database^\Question$. \\ \cline{2-5} 
 & \textsf{FILTER}    & $\sigma_{\textit{cond}}(\Instance)$ & \textit{``Return rows from $\Instance$ where $A$ [Op] $V$.''} & Select tuples $\Tuple \in \Instance$ where $A \in \AttrS$ meets a condition. \\ \cline{2-5} 
 & \textsf{PROJECT}   & $\pi_{A_1, \dots, A_k}(\Instance)$ & \textit{``Return $\{A_1, \dots, A_k\}$ of $\Instance$.''} & Projection over a subset of attributes $\{A_j\} \subseteq \AttrRel$. \\ \cline{2-5} 
 & \textsf{AGGREGATE} & $\gamma_{\textit{f}(A), \textit{G}}(\Instance)$ & \textit{``Return $\textit{f}(A)$ grouped by $\textit{G}$ from $\Instance$.''} & Deterministic reductions for $A \in \AttrS$ grouped by $\textit{G}$. \\ \cline{2-5} 
 & \textsf{JOIN}      & $\Instance_a \bowtie_{\theta} \Instance_b$ & \textit{``Return combined rows from $\Instance_a, \Instance_b$ via $\theta$.''} & Join operation based on exact key matching $\theta$. \\ \cline{2-5} 
 & \textsf{SORT}      & $\tau_{A}(\Instance)$ & \textit{``Return $\Instance$ sorted by $A$.''} & Reorder tuples $\Tuple \in \Instance$ by scalar values of attribute $A \in \AttrS$. \\ \cline{2-5} 
 & \textsf{LIMIT}     & $\lambda_{n}(\Instance)$ & \textit{``Return the top $n$ rows from $\Instance$.''} & Truncate the instance to the first $n$ tuples. \\ \cline{2-5} 
 & \textsf{SET\_OP}   & $\Instance_a \{ \cup, \cap, \setminus \} \Instance_b$ & \textit{``Return the [Op] of $\Instance_a$ and $\Instance_b$.''} & Standard set operations between compatible relations. \\ \hline
\multirow{4}{*}{\rotatebox[origin=c]{90}{\textbf{Semantic}}} 
 & \textsf{MAP} & $\tilde{\pi}_{\textit{cond}}(\Instance)$ & \textit{``Return $\Instance$ with new col derived from $\AttrU$ by \textit{cond}.''} & Row-wise transform (e.g., sentiment or entity extraction from $A \in \AttrU$). \\ \cline{2-5} 
 & \textsf{FILTER} & $\tilde{\sigma}_{\textit{cond}}(\Instance)$ & \textit{``Return rows from $\Instance$ satisfying \textit{cond}.''} & Probabilistic tuple selection based on fuzzy NL intent. \\ \cline{2-5} 
 & \textsf{JOIN}   & $\Instance_a \tilde{\bowtie}_{\textit{cond}} \Instance_b$ & \textit{``Join $\Instance_a, \Instance_b$ via matching logic: \textit{cond}.''} & Fuzzy entity resolution across relations via semantic similarity. \\ \cline{2-5} 
 & \textsf{AGGREGATE}    & $\tilde{\gamma}_{\textit{f}(A), \textit{G}}(\Instance)$ & \textit{``Return summary of $\AttrU$ grouped by $\textit{G}$ from $\Instance$ via \textit{cond}.''} & Synthesis or summarization of textual content in $\AttrU$. \\ \bottomrule
\end{tabular}
\caption{\small The \System Operator Universe}
\label{tab:operators}
\end{table*}

\subsection{PLANNING: Deterministic Optimization of Query Plans}
\label{app:sec:plan_optimization_details}
\begin{algorithm}[!tb]
\small
\caption{Heuristic-Based Plan Optimization of \System} \label{alg:optimizer}
\begin{algorithmic}[1]
    \Require Plan $\mathcal{G}$, Thresholds $\varepsilon, \tau$
    \Ensure Optimized Plan $\mathcal{G}^*$
    
    \Function{\textsc{OptimizePlan}}{$\mathcal{G}, \varepsilon, \tau$}
        \State $\mathcal{G}_{pushed} \gets \textsc{PushSelections}(\mathcal{G})$
        \State $\mathcal{G}_{pruned} \gets \textsc{PruneProjections}(\mathcal{G}_{pushed})$
        
        \State $\mathcal{G}_{ordered} \gets \textsc{ReorderJoins}(\mathcal{G}_{pruned}, \tau)$
        
        \For{each semantic node $v_{sem} \in \mathcal{G}$}
            \State $v_{down} \gets \textsc{GetNext}(v_{sem})$
            
            \If{$v_{down} \text{ is a \textsf{JOIN} } A \bowtie_{k} B$}
                \State $\gamma_{out} \gets (|A| \cdot |B|) / \max(| \pi_{key}(A) |, | \pi_{key}(B) |)$
            \ElsIf{$v_{down} \text{ is a \textsf{UNION} } A \cup B$}
                \State $\gamma_{out} \gets |A| + |B|$
            \EndIf
            
            \If{$v_{down} \text{ is a \textsf{JOIN} or \textsf{UNION}}$}
                \State $\Delta \gamma \gets \gamma_{out} / \gamma_{in}(v_{sem})$
                \If{$\Delta \gamma > \varepsilon$}
                    \State $\mathcal{G} \gets \textsc{Elevate}(v_{sem}, v_{down})$
                \Else
                    \State $\mathcal{G} \gets \textsc{Defer}(v_{sem})$
                \EndIf
            \EndIf
        \EndFor
        
        \State $\mathcal{G}^* \gets \mathcal{G}_{ordered}$
        \State \Return $\mathcal{G}^*$
    \EndFunction
    \Statex

    \Function{\textsc{ReorderJoins}}{$\mathcal{G}, \tau$}
        \State $\mathcal{R} \gets \mathcal{G}.\Schema$
        
        \If{$|\mathcal{R}| \le \tau$}
            \State $\mathcal{G}.\text{join\_tree} \gets \textsc{DP}(\mathcal{R})$ 
        \Else
            \State $\mathcal{G}.\text{join\_tree} \gets \textsc{Greedy}(\mathcal{R})$ 
        \EndIf
        
        \State \Return $\mathcal{G}$
    \EndFunction
    \Statex
    
    \Function{\textsc{Greedy}}{$\mathcal{R}$}
        \State $(\Relation_i, \Relation_j) \gets \arg\min_{\Relation_x, \Relation_y \in \mathcal{R}} \textsc{Cost}(\Relation_x \bowtie \Relation_y)$
        \State $Plan \gets \Relation_i \bowtie \Relation_j; \quad \mathcal{R} \gets \mathcal{R} \setminus \{\Relation_i, \Relation_j\}$
        
        \While{$\mathcal{R} \neq \emptyset$}
            \State $\Relation_{next} \gets \arg\min_{\Relation_k \in \mathcal{R}} \textsc{Cost}(Plan \bowtie \Relation_k)$
            \State $Plan \gets Plan \bowtie \Relation_{next}; \quad \mathcal{R} \gets \mathcal{R} \setminus \{\Relation_{next}\}$
        \EndWhile
        
        \State \Return $Plan$
    \EndFunction
    \Statex
    
    \Function{\textsc{DP}}{$\mathcal{R}$}
        \State $Opt \gets \emptyset$
        \For{each $\Relation_i \in \mathcal{R}$} 
            \State $Opt[\{\Relation_i\}] \gets \Relation_i$ 
        \EndFor
        
        \For{$size \gets 2$ \textbf{to} $|\mathcal{R}|$}
            \For{each subset $S \subseteq \mathcal{R}$ \textbf{where} $|S| = size$}
                \State $Opt[S] \gets \arg\min_{\Relation \in S} \textsc{Cost}(Opt[S \setminus \{\Relation\}] \bowtie \Relation)$
            \EndFor
        \EndFor
        
        \State \Return $Opt[\mathcal{R}]$
    \EndFunction
    \Statex
    
    \Function{\textsc{Cost}}{$A \bowtie B$}
        \State $\gamma_{in} \gets |A| + |B|$;\quad $pages \gets \textsc{EstimatePages}(A, B)$
        
        \State \Return $\gamma_{in} \cdot c_{cpu} + pages \cdot c_{io}$
    \EndFunction
\end{algorithmic}
\end{algorithm}

{As previously mentioned in Section~\ref{ssec:planning}, the execution plan generated by the planner may not be cost-efficient, particularly when semantic operators are involved. Aside from the LLM-instructed approach discussed in Section~\ref{ssec:planning},\System also extends classical relational query optimization techniques ~\cite{selinger1979access, hellerstein1993predicate} to account for the asymmetry between inexpensive relational operations and expensive semantic reasoning in a deterministic manner. Specifically, we adopt predicate migration as a \textit{cost-aware semantic placement} strategy. While semantic operators are deferred to later stages to minimize unnecessary LLM calls, the optimizer evaluates the \textit{cardinality impact} of downstream relational operators. If a join or set operation is expected to significantly increase the number of tuples, semantic filters are applied earlier on smaller base relations to avoid costly expansion. The optimizer applies the following transformations:}

\begin{itemize}[leftmargin=*]
    \item \textbf{Selection Pushing:} 
    {When possible, relational filters are pushed toward the base relations to eliminate irrelevant tuples early in the plan.}
    
    \item \textbf{Projection Pruning:} 
    {Unnecessary attributes are removed once they are no longer required by downstream operations. This reduces the width of intermediate relations and memory footprint.}
    
    \item \textbf{Join Reordering:} 
    {Join operations are reordered to prioritize high-selectivity joins and minimize intermediate relation sizes.}
    
    \item \textbf{Adaptive Semantic Deferral:} Semantic steps are deferred until relational pruning is complete, unless a relational operator threatens to expand the result set, in which case the semantic step is prioritized to act as a pre-filter.
\end{itemize}

To formalize these heuristics, we define our join reordering strategy within the optimization framework shown in Algorithm~\ref{alg:optimizer}. For join reordering (\textsc{ReorderJoins}), we employ a threshold-based hybrid strategy that balances optimality and scalability. When the number of base relations is small ($|\mathcal{R}| \le \tau$), we use dynamic programming to search the space of left-deep join trees and obtain the cost-optimal plan. For larger queries, we fall back to a greedy heuristic that iteratively selects the relation minimizing intermediate join cost.
To guide semantic operator placement, we define a cost model that captures both relational execution cost and LLM inference cost. For a query plan DAG $\mathcal{G}$, the total estimated execution cost is:
$$Cost(\mathcal{G})=\sum_{o \in \mathcal{G}}\Big(w_{sys} \cdot C_{sys}(o)+w_{llm} \cdot C_{llm}(o)\Big)$$
where $w_{sys}$ and $w_{llm}$ are weighting factors. For relational operators, $C_{llm}(o)=0$, and the system cost $C_{sys}(o)$ is defined by traditional I/O and CPU estimates:
$$C_{sys}(o)=\gamma_{in}(o) \cdot c_{cpu}+pages(o) \cdot c_{io}$$
where $\gamma_{in}(o)$ denotes the estimated input cardinality for operator $o$. For a set operation like UNION ($A \cup B$), the output cardinality $\gamma_{out}$ is simply the additive sum of its inputs ($|A| + |B|$). For a join operator $A \bowtie B$, the input cardinality is the sum of the sizes of the two relations ($|A| + |B|$), while the output cardinality $\gamma_{out}$ (which becomes the input $\gamma_{in}$ for subsequent operators) is estimated using the selectivity of the join key:
$$\gamma_{out} = \frac{|A| \cdot |B|}{\max(| \pi_{key}(A) |, | \pi_{key}(B) |)}$$ 
where $| \pi_{key}(R) |$ represents the number of distinct values of the join attribute in relation $R$. This value is typically maintained in the system catalog, estimated via HyperLogLog sketches during data ingestion or periodic statistics updates~\cite{chen2017two, muller2021memory}. The parameters $c_{cpu}$ is the calibrated unit costs for CPU cycles per tuple,  $c_{io}$ is I/O disk page fetches and $pages(o)$ is total data volume in memory pages required for the operation.

For semantic operators, the dominant cost arises from LLM inference. We model this as:
$$C_{llm}(o)=\gamma_{in}(o) \cdot \Big(c_{call}+\alpha \cdot \mathbb{E}[|tokens(A)|]\Big)$$
where $c_{call}$ denotes the base API latency, $\alpha$ is the marginal processing cost per token, and $\mathbb{E}[|tokens(A)|]$ is the expected token length of the evaluated attributes. This formulation motivates \textit{Adaptive Semantic Deferral}: minimizing the input cardinality $\gamma_{in}(o)$ through relational pruning directly reduces the LLM inference cost $C_{llm}(o)$. 

Rather than computing exact costs for every placement, the optimizer relies on a calibrated threshold $\varepsilon$ that captures the break-even point between relational data expansion and expensive LLM inference. It further considers the cardinality impact of downstream operators. Let $\Delta\gamma=\gamma_{out}/\gamma_{in}$ denote the ratio between the estimated output cardinality of a downstream operator and the input cardinality of the semantic branch. If $\Delta \gamma > \varepsilon$ (where $\varepsilon \ge 1$), the downstream join or set operation acts as a data multiplier. In such cases, executing the semantic filter earlier on the smaller base relation prevents a multiplicative increase in LLM token consumption.

The cost model parameters are obtained through a lightweight calibration phase. Relational parameters ($c_{cpu},c_{io}$) are estimated using synthetic workloads, while semantic parameters ($c_{call},\alpha$) are measured from LLM API latency and token statistics. Expected token lengths are maintained in the system metadata catalog.

{In summary, Algorithm~\ref{alg:optimizer} rewrites an initial logical plan into a cost-optimized physical plan. It first applies rule-based transformations such as filter and projection pushdown, then performs hybrid join reordering, and finally determines semantic operator placement using the cost-aware deferral heuristic.}

{While our cost model provides a practical approximation of hybrid execution cost, it is not intended to be fully precise. Instead, it serves as a lightweight decision mechanism to guide operator placement and plan optimization. Empirically, we observe that even coarse-grained estimates are sufficient to achieve significant cost reductions, suggesting that precise modeling of LLM behavior may not be necessary for effective optimization.}

\subsection{EXECUTION: Batch Execution of Semantic Operators}
\label{app:sec:batch_execution_details}
Algorithm~\ref{alg:llm_batching} outlines the complete batched execution logic employed by the LLM-based reasoning engine of \System.

\begin{algorithm}[!t]
\small
\caption{Batching Semantic Operators} \label{alg:llm_batching}
\begin{algorithmic}[1]
    \Require Operator $Op$, Instruction $I$, Input Relation $\Instance$, Token Budget $B_{max}$, Base Batch Size $b$, Token Cost per Row $t_{row}$
    \Ensure Result Relation $\Instance_{out}$
    
    \Function{\textsc{ExecuteSemantic}}{$Op, I, \Instance, B_{max}, b, t_{row}$}
        \State $\beta \gets \min\Big(b, \lfloor B_{max} / t_{row} \rfloor \Big)$; \quad
        $\Instance_{out} \gets \emptyset$
        
        \If{$Op = \textsf{MAP}$}
            \For{$i \gets 0$ \textbf{to} $|\Instance|-1$ \textbf{step} $\beta$}
                \State $C \gets \Instance[i : i+\beta]$; \quad
                $Col_{new} \gets \textsc{Map}(I, C)$
                \State $\Instance_{out} \gets \Instance_{out} \cup \{ C[j] \oplus Col_{new}[j] \mid j \in \{0, \dots, |C|-1\} \}$
            \EndFor
            
        \ElsIf{$Op = \textsf{FILTER}$}
            \For{$i \gets 0$ \textbf{to} $|\Instance|-1$ \textbf{step} $\beta$}
                \State $C \gets \Instance[i : i+\beta]$ ;\quad
                 $Indices \gets \textsc{Filter}(I, C)$
                \State $\Instance_{out} \gets \Instance_{out} \cup \{ C[j] \mid j \in Indices \}$
            \EndFor
            
        \ElsIf{$Op = \textsf{JOIN}$}
            \For{$i \gets 0$ \textbf{to} $|\Instance|-1$ \textbf{step} $\beta$}
                \State $C_A \gets \Instance[i : i+\beta]$
                \For{$j \gets 0$ \textbf{to} $|\Instance_B|-1$ \textbf{step} $\beta$}
                    \State $C_B \gets \Instance_B[j : j+\beta]$;\quad
                     $res \gets \textsc{Join}(I, C_A, C_B)$
                    \State $\Instance_{out} \gets \Instance_{out} \cup res$
                \EndFor
            \EndFor
            
        \ElsIf{$Op = \textsf{AGGREGATE}$}
            \State $\Instance_{out} \gets \textsc{Reduce}(I, \Instance, \beta, 0)$
        \EndIf
        
        \State \Return $\Instance_{out}$
    \EndFunction
    \Statex
    
    \Function{\textsc{Reduce}}{$I, \Instance, \beta, depth$}
        \If{$|\Instance| \le \beta$ \textbf{or} $depth > MAX\_DEPTH$}
            \State \Return $\textsc{Aggregate}(I, \Instance)$
        \EndIf
        
        \State $\Instance_{partial} \gets \emptyset$
        \For{$i \gets 0$ \textbf{to} $|\Instance|-1$ \textbf{step} $\beta$}
            \State $C \gets \Instance[i : i+\beta]$;\quad
             $res \gets \textsc{Aggregate}(I, C)$
            \State $\Instance_{partial} \gets \Instance_{partial} \cup res$
        \EndFor
        
        \State \Return $\textsc{Reduce}(I, \Instance_{partial}, \beta, depth + 1)$
    \EndFunction
\end{algorithmic}
\end{algorithm}

\section{Experimental Settings}
\label{app:sec:experimental_details}
We provide detailed information about the datasets, base models, and prompts used in our experiments.

\subsection{Dataset Details}
\label{app:sec:dataset_details}
For datasets containing unstructured text (\textsc{FeTaQA}, \textsc{HybridQA}, \textsc{TAT-QA}), we convert them into a semi-structured format by appending the associated text as an additional column to each row.
We evaluate on 100 samples per dataset, except for \textsc{RepairTQA}-\textsc{M2}, where we use 50 samples. To manage cost, we run all effectiveness experiments using \textsc{Gemini-3-Flash-Preview}, and evaluate additional models on a challenging subset of \textsc{RepairTQA}. All datasets include gold-standard annotations used for evaluation.

The details of other semi-structured datasets are as follows:
\textsc{FeTaQA} \cite{nan2022fetaqa}, a free-form table QA dataset that requires generating descriptive answers from retrieved information; \textsc{HybridQA} \cite{chen2020hybridqa}, a multi-hop QA dataset that requires integrating information from structured tables and linked unstructured text; and \textsc{TAT-QA} \cite{zhu2021tatqa}, a financial-domain dataset requiring complex numerical reasoning over hybrid contexts of tables and text.

\subsection{Base Model Details}
\label{app:sec:models}
Table \ref{tab:models} summarizes the base models used in our experiments, including their context window sizes and licensing information.

\begin{table*}[t]
  \centering
  \resizebox{\linewidth}{!}{
  \begin{tabular}{lrrll}
      \toprule
      \textbf{Model} & \textbf{Size} & \textbf{Context} & \textbf{HuggingFace / API} & \textbf{License}\\
      \midrule
      GPT-5-mini \citep{openai2025gpt5} & --- & 400k & \texttt{gpt-5-mini-2025-08-07} & OpenAI Service Terms\\
      Gemini-3-Flash \citep{google2025gemini3flash} & --- & 1M & \texttt{gemini-3-flash-preview} & Gemini API Additional Terms of Service\\
      Qwen-3 \citep{yang2025qwen3} & 30B & 128k & \texttt{qwen3:30B} & Apache license 2.0 \\
      \bottomrule
  \end{tabular}
  }
  \caption{Base models used in experiments. Model sizes are not publicly disclosed (---).}
  \label{tab:models}
\end{table*}

\section{Extended Experimental Evaluations}
\label{app:sec:extended_exp}
We present additional experimental results that complement the main findings in the paper, including hyperparameter analysis, extended effectiveness comparisons across multiple models, and an ablation study.

\subsection{Evaluating Conventional TQA Benchmarks}
\label{app:sec:conventional_tqa}
Although the majority of our experiments focus on more challenging, semi-structured TQA datasets, we also evaluate our approach on conventional TQA benchmarks, namely \textsc{WikiTableQuestions} and \textsc{WikiSQL}. As shown in Figure~\ref{fig:acc_gemini_traditional}, the results are consistent with our observations on other datasets, with \System consistently emerging either as the top-performing method or a close runner-up. 

\subsection{Evaluating Additional Plan Consolidation Strategies}
\label{app:sec:alt_consolidation}
\begin{figure*}[t] 
\centering
    \includegraphics[width=0.9\textwidth]{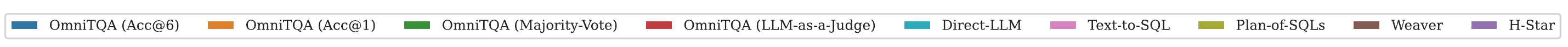}\\
    \begin{minipage}[t]{0.22\linewidth}
        \centering
        \includegraphics[width=\textwidth]{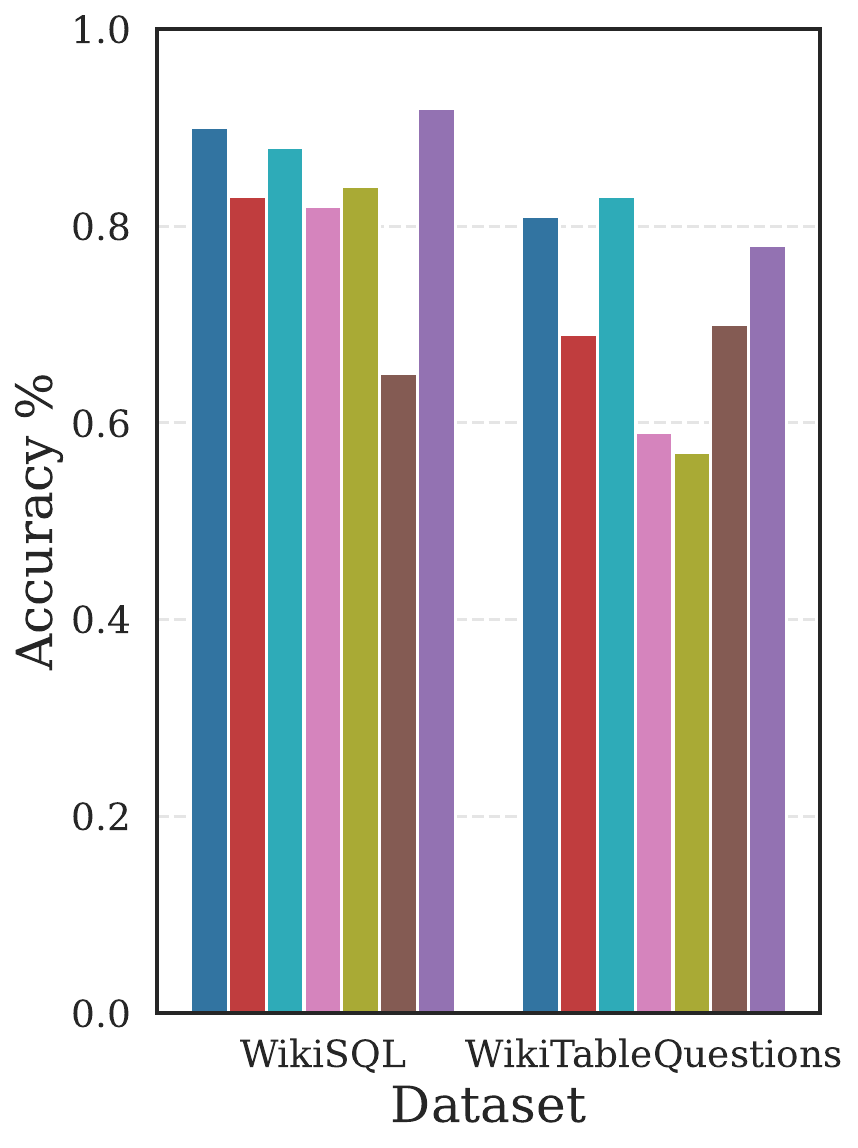}
        \vspace{-2.0em}
        \caption{\small Effectiveness comparison of \System vs. baselines on conventional TQA benchmarks evaluated with \textsc{Gemini-3-Flash}.}
        \label{fig:acc_gemini_traditional}
    \end{minipage}
    \begin{minipage}[t]{0.32\linewidth}
        \centering
        \includegraphics[width=\textwidth]{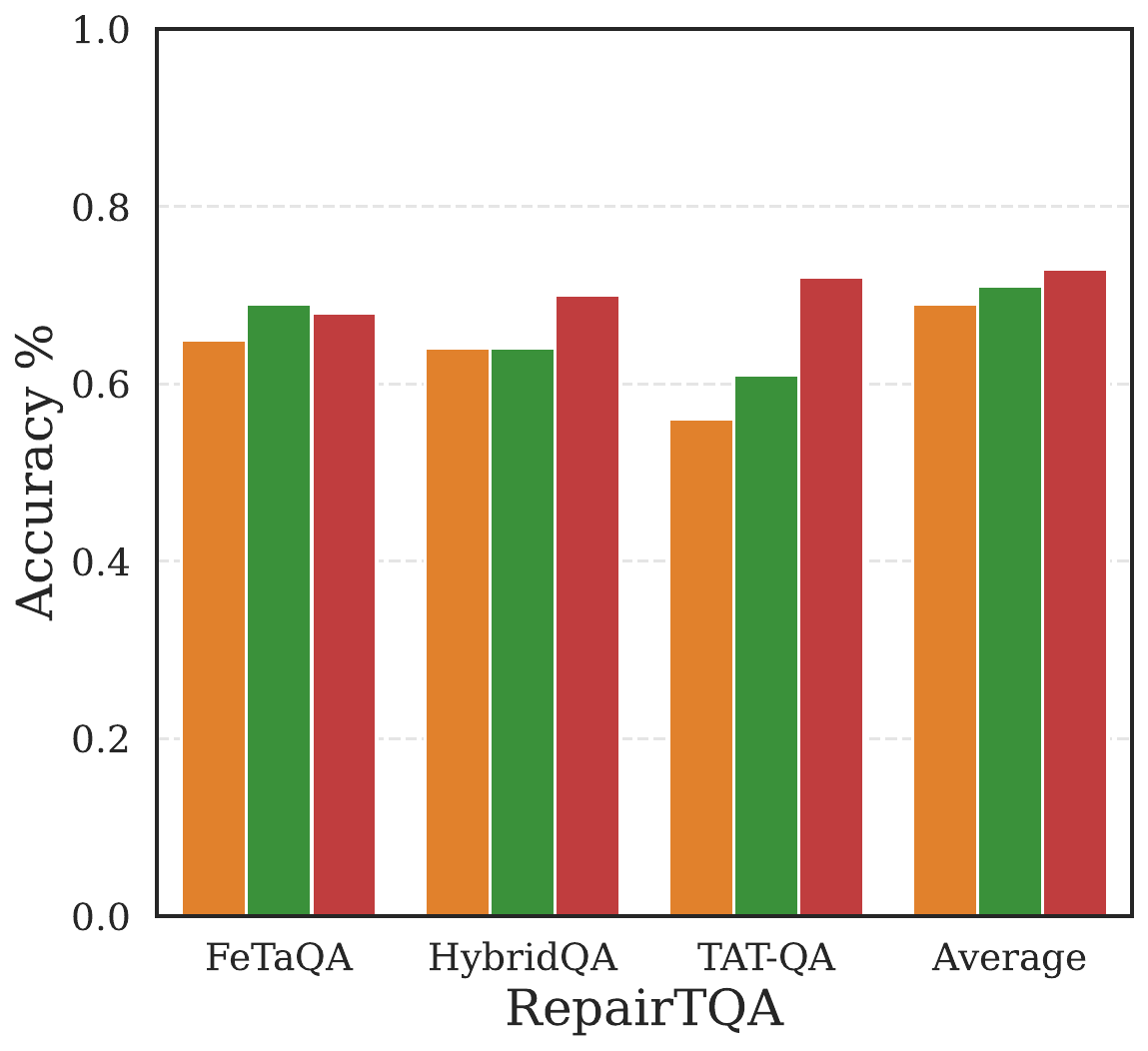}
        \vspace{-2.0em}
        \caption{\small Extended answer consolidation results in \System using \textsc{Gemini-3-Flash}.}
        \label{fig:acc_gemini_agg}
    \end{minipage}
    \begin{minipage}[t]{0.20\linewidth}
        \centering
        \includegraphics[width=\textwidth]{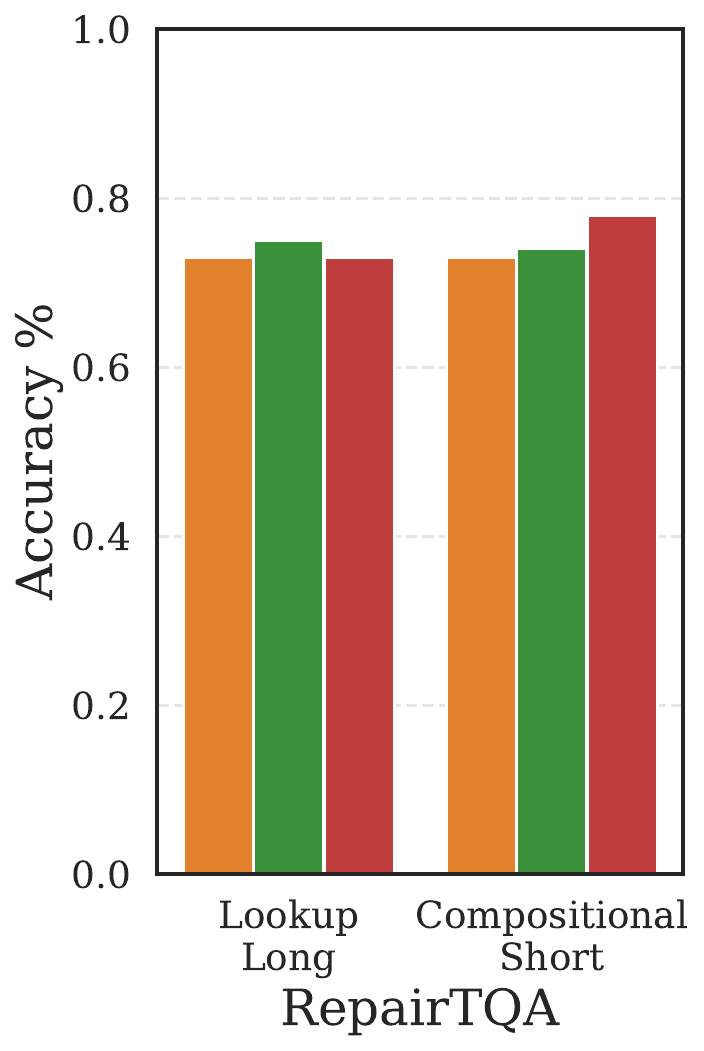}
        \vspace{-2.0em}
        \caption{\small Extended answer consolidation results in \System using \textsc{Gpt-5-mini}.}
        \label{fig:acc_gpt_agg}
    \end{minipage}
    \begin{minipage}[t]{0.20\linewidth}
        \centering
        \includegraphics[width=\textwidth]{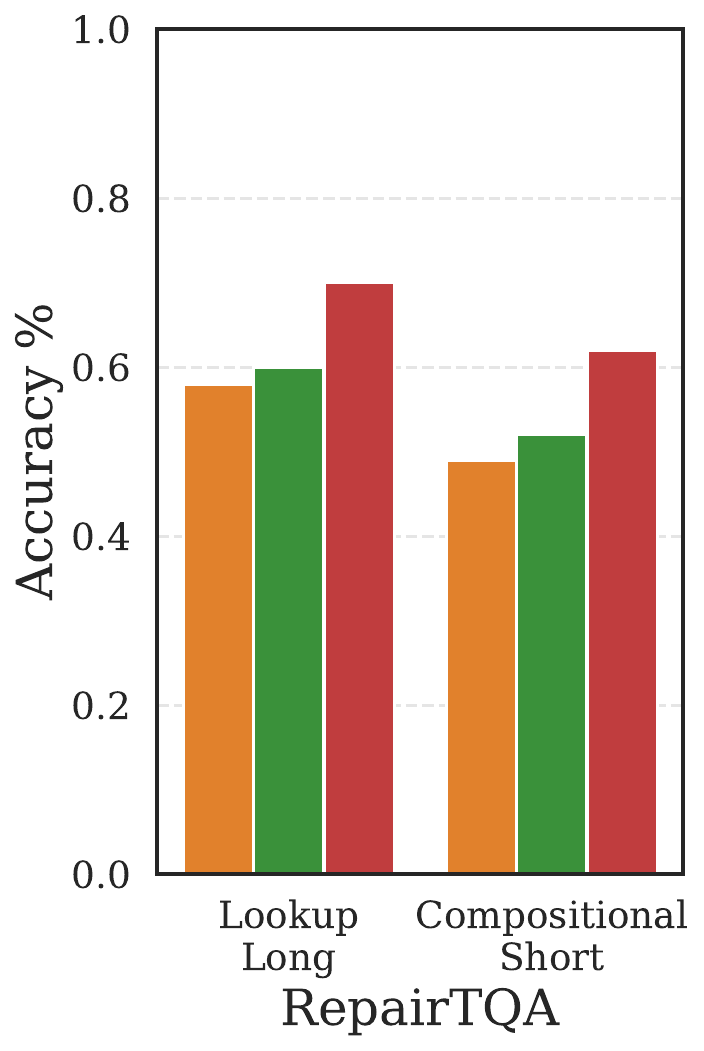}
        \vspace{-2.0em}
        \caption{\small Extended answer consolidation results in \System using \textsc{Qwen3:30B}.}
        \label{fig:acc_qwen_agg}
    \end{minipage}
\end{figure*}

In addition to the \textsc{Acc@6} and LLM-as-a-Judge metrics presented in the main text, we also evaluate our system using \textsc{Acc@1} (single-plan generation) and Semantic Majority Vote strategies. Figures~\ref{fig:acc_gemini_agg},~\ref{fig:acc_gpt_agg}, and~\ref{fig:acc_qwen_agg} illustrate these results across various base LLMs. The observed trends align with our primary findings, with the LLM-as-a-Judge strategy outperforming these alternative approaches in the majority of configurations.

\subsection{Hyperparameter Analysis}
\label{app:sec:hyperparameter_analysis}
We examine how hyperparameters influence the effectiveness and efficiency of \System. For these and all following experiments, we employ \textsc{Gemini-3-Flash-Preview} as the base model.

\begin{figure*}[t]
\centering
    \begin{minipage}[t]{0.24\linewidth}
        \centering
        \includegraphics[width=\textwidth]{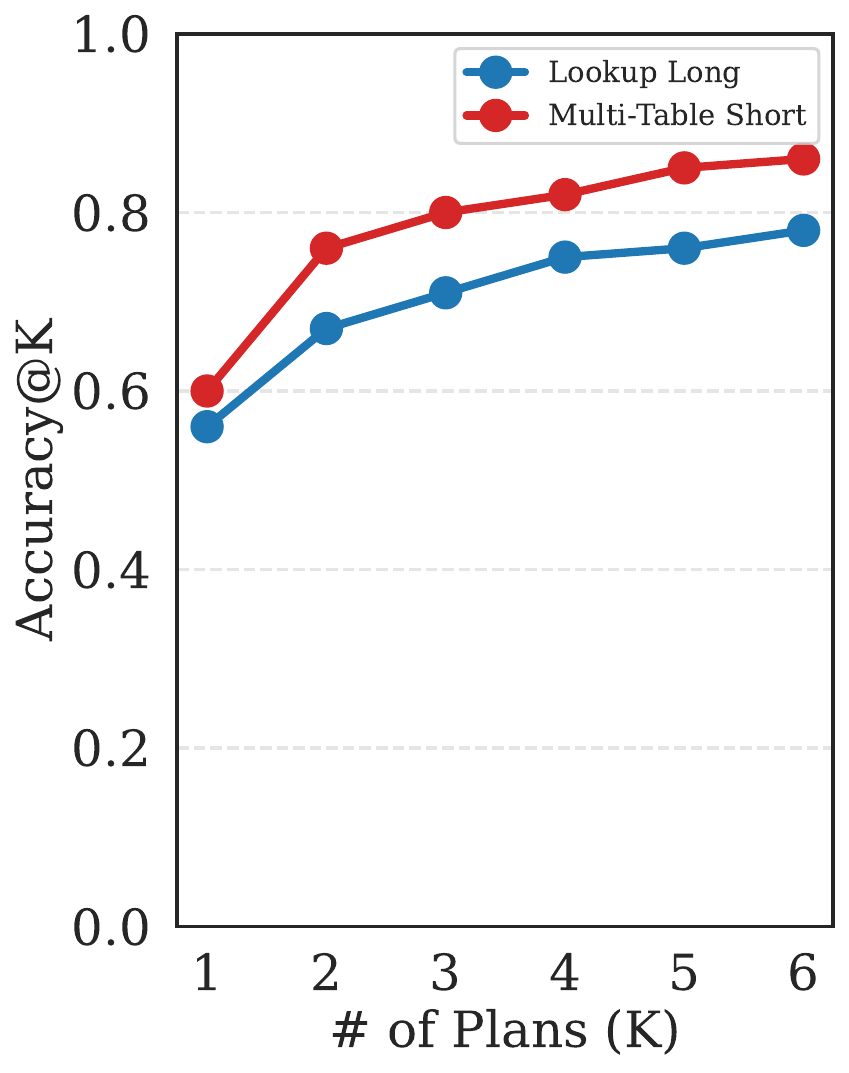}
        \vspace{-2.0em}
        \caption{\small Varying no. of plans $K$ vs. accuracy.}
        \label{fig:varying_k_acc}
    \end{minipage}
    \hfill
    \begin{minipage}[t]{0.24\linewidth}
        \centering
        \includegraphics[width=\textwidth]{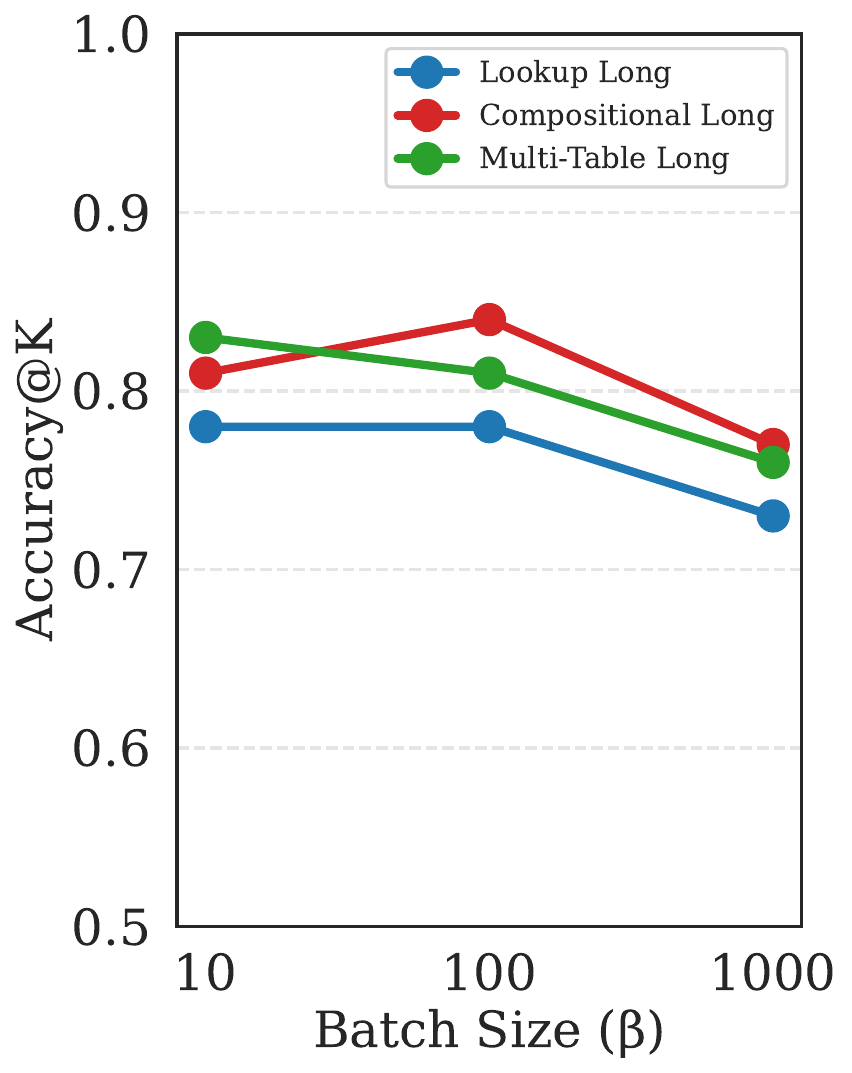}
        \vspace{-2.0em}
        \caption{\small Varying batch size $\beta$ vs. accuracy.}
        \label{fig:varying_beta_acc}
    \end{minipage}
    \hfill
    \begin{minipage}[t]{0.24\linewidth}
        \centering
        \includegraphics[width=\textwidth]{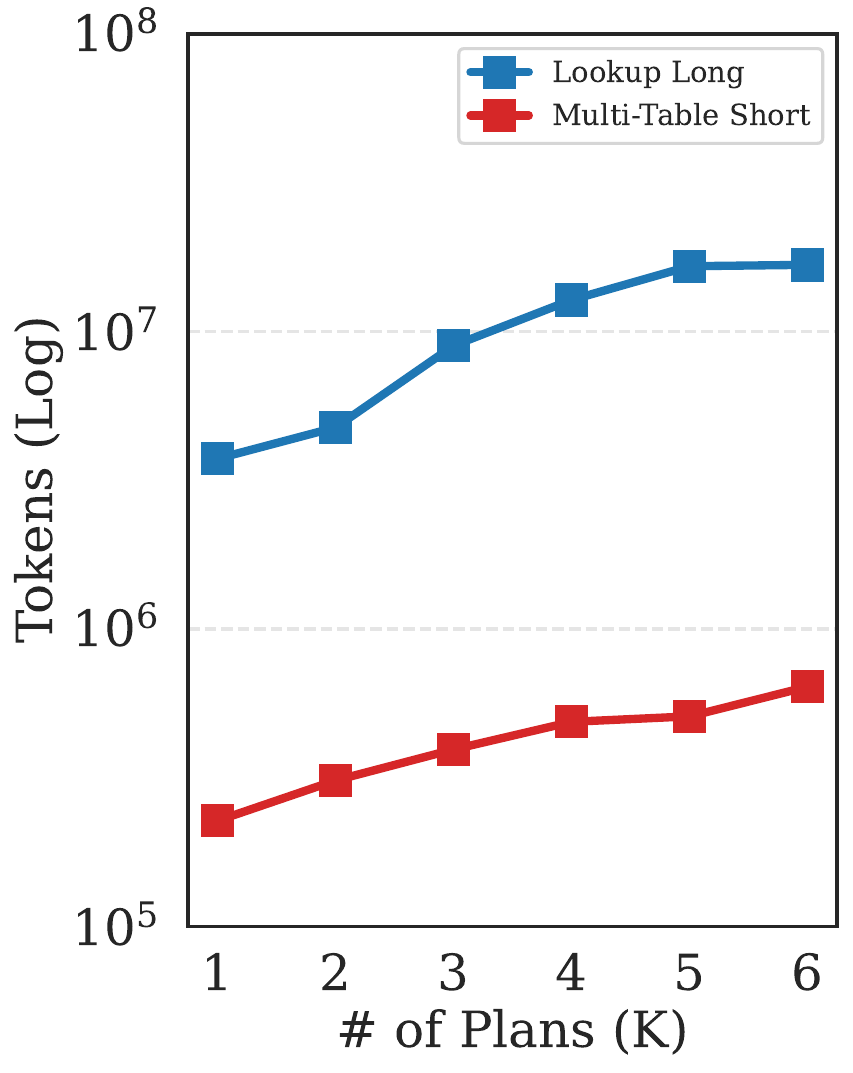}
        \vspace{-2.0em}
        \caption{\small Varying no. of plans $K$ vs. cost.}
        \label{fig:varying_k_cost}
    \end{minipage}
    \hfill
    \begin{minipage}[t]{0.24\linewidth}
        \centering
        \includegraphics[width=\textwidth]{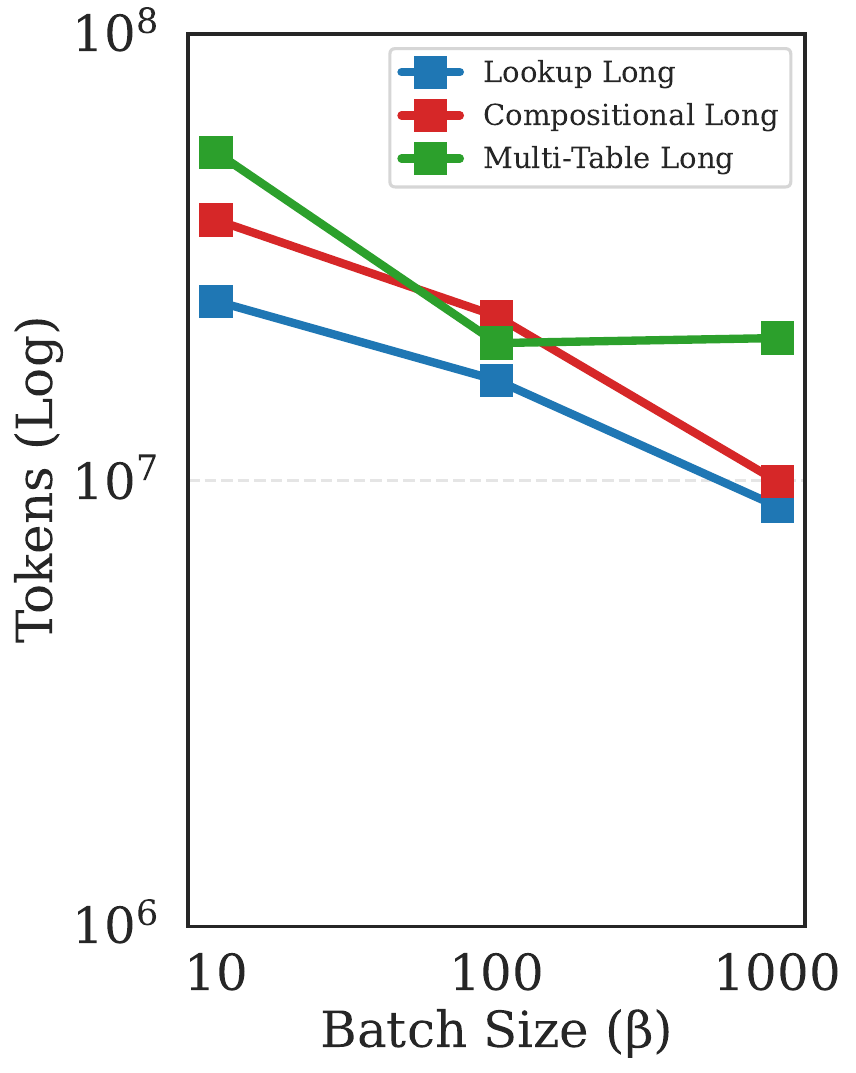}
        \vspace{-2.0em}
        \caption{\small Varying batch size $\beta$ vs. cost.}
        \label{fig:varying_beta_cost}
    \end{minipage}
\end{figure*}

\stitle{Number of Plans ($K$)}
We analyze the effect of varying the number of plans ($K$) on accuracy (Figure~\ref{fig:varying_k_acc}). We observed that accuracy improves as $K$ increases, with the largest gain observed from $K=1$ to $K=2$, after which performance plateaus. This suggests that a small number of plans already captures most of the benefit, and further increases yield diminishing returns relative to cost.

In terms of cost, as shown in Figure~\ref{fig:varying_k_cost}, total cost scales linearly with $K$, since each additional plan incurs a full generation and execution cycle.

\stitle{Batch Size ($\beta$)}
Next, we evaluate the impact of varying the batch size ($\beta$) on the accuracy of \System. As illustrated in Figure~\ref{fig:varying_beta_acc}, accuracy generally decreases as $\beta$ increases, with more pronounced degradation at larger values, likely due to longer contexts introducing noise or hallucinations. These results highlight the importance of empirically tuning $\beta$; such optimization could be performed via expert manual intervention or through a small-scale validation (e.g., binary search over a representative subset).

Regarding the impact on cost, Figure~\ref{fig:varying_beta_cost} shows that cost decreases with larger $\beta$, due to more efficient batching that reduces the number of LLM calls and eliminates the redundant processing of shared prompt overhead.

\subsection{Additional Ablation Study Results}
\label{app:sec:additional_ablation}
We also repeated our ablation study using the \textsc{Acc@6} metric, reporting the outcomes in Table~\ref{tab:ablation_acc6_appendix}. These findings are entirely consistent with the LLM-as-a-Judge evaluation discussed in the main body.

\providecommand{\vhead}[1]{\rotatebox[origin=c]{90}{\textbf{#1}}}

\begin{table}[t]
\scriptsize
\centering
\vspace{-1em}
{\setlength{\tabcolsep}{2.5pt}
\resizebox{\columnwidth}{!}{
\begin{tabular}{l|c|ccccc}
    \toprule
    \textbf{Dataset} & \vhead{\System} & \vhead{w/ Naive $\Preview$} & \vhead{w/ QDMR} & \vhead{w/o Opt. $\mathcal{G}$} & \vhead{w/o Div. $\mathcal{G}$} & \vhead{w/o Pru. $\Schema$} \\ \midrule
Lookup Long & \textbf{0.78} & \cellcolor{moddrop}0.70 & \cellcolor{sigdrop}0.53 & 0.78 & \cellcolor{moddrop}0.66 & \cellcolor{slidrop}0.77 \\ \hline
Composional Short & \textbf{0.86} & 0.86 & \cellcolor{sigdrop}0.57 & 0.86 & \cellcolor{moddrop}0.79 & \cellcolor{slidrop}0.83 \\ \hline
Multi-Table Short & \textbf{0.86} & \cellcolor{moddrop}0.78 & \cellcolor{sigdrop}0.37 & 0.86 & \cellcolor{moddrop}0.72 & \cellcolor{slidrop}0.82 \\
\bottomrule
\end{tabular}
}}
\caption{Ablation study: accuracy under \textsc{Acc@6}.}
\label{tab:ablation_acc6_appendix}
\end{table}

\subsection{Manual Inspection for Failure Cases on \textsc{RepairTQA}}
\label{app:sec:failure_cases}
A manual audit of \System’s performance on \textsf{RepairTQA} reveals that most failures are approach-agnostic and stem from dataset limitations, including incorrect labels, ambiguous questions, and structural issues such as missing or incomplete data. Additionally, the semantic evaluation protocol occasionally produces false negatives. While such benchmark noise is well documented~\cite{wretblad-etal-2024-understanding}, we retain these cases to ensure a consistent and unbiased comparison. Excluding them would uniformly inflate performance across all methods.

\subsection{Additional Error Analysis}
\label{app:sec:error_analysis_details}

{\stitle{Schema grounding errors} These occur when query terms are incorrectly mapped to schema attributes. For example, similarity between column names may lead to selecting \textsf{ProductCategoryID} instead of \textsf{ProductSubcategoryID}. In other cases, the system struggles with implicit semantics, such as mapping `has a text box' to a negated boolean field (\textsf{isTextLess}). Such schema grounding errors are well-known in Text-to-SQL systems, and can be sometimes resolved with sophisticated schema linking.}

{\stitle{Execution errors} Execution errors are less frequent and typically involve semantic operators. These include incorrect attribute extraction despite correct row retrieval (e.g., returning `province' instead of `country', as well as incomplete retrieval where valid rows are missed.}

{\stitle{Robustness to data vs. prior knowledge} We further observe that the execution engine prioritizes database content over parametric knowledge. To evaluate this robustness, we performed a counterfactual analysis by altering the values associated with specific filters in several samples. For instance, in the query \textit{``List all cities in the Northern California Region,''} we replaced the relevant records with cities from New York. The the system faithfully returns modified data, indicating strong grounding to the input rather than memorized world knowledge.}

\clearpage
\onecolumn
\section{Prompts}
\label{app:sec:prompts}
\begin{flushleft}
In the following, we provide all the prompts used in our experiments in different steps of \System:
\end{flushleft}

\newcommand{\promptcaption}[2]{%
    \captionsetup{type=figure,justification=raggedright,singlelinecheck=false}%
    \captionof*{figure}{\textbf{#1}: #2}%
    \vspace{-0.75em}%
}

\promptcaption{Preprocessing Prompt}{Semantic Schema Pruning}
\begin{tcolorbox}[
    title=Semantic Schema Pruning: Prompt,
    breakable, 
    sharp corners,
    colback=gray!3,
    colframe=gray!75,
    fonttitle=\bfseries,
    listing only,  
    listing options={
        basicstyle=\ttfamily\footnotesize,
        breaklines=true,
        columns=fullflexible,
        keepspaces=true
    }
]
    \footnotesize
\begin{lstlisting}
You are an expert in Text-to-SQL pipelines. Your specific task is "Schema Pruning": filtering a database schema to a subset of columns relevant to a natural language question. Your Goal is to Maximize Recall. It is critical that you include ALL columns that might possibly be needed to answer the question, including columns for filtering, joining, grouping, or sorting.

**The "High-Recall" Protocol:**
1. If a column is a Primary Key or Foreign Key, KEEP IT.
2. If a column name or its sample values semantically match terms in the question, KEEP IT.
3. If the question implies a time frame (e.g., "recent", "trend", "when"), KEEP date/timestamp columns.
4. If you are 50/50 split on whether a column is relevant, KEEP IT.
5. Only exclude a column if you are certain it is noise.

### TABLE ###
{table}

### QUESTION ### 
{question}

### Available Columns ### 
(Name (Type): [Samples]): {context_str}

Task: Return a JSON list of strings containing the columns relevant to the question according to the High-Recall Protocol.
Output strictly valid JSON.
\end{lstlisting}
\end{tcolorbox}


\promptcaption{Planning Prompt}{Decomposition: System Prompt}

\begin{tcolorbox}[
    title=Decomposition: System Prompt,
    breakable, 
    sharp corners,
    colback=gray!3,
    colframe=gray!75,
    fonttitle=\bfseries,
    listing only,  
    listing options={
        basicstyle=\ttfamily\footnotesize,
        breaklines=true,
        columns=fullflexible,
        keepspaces=true
    }
]
    \footnotesize
\begin{lstlisting}
You are a query planner specializing in question decomposition. Your goal is to decompose a natural language question into **up to {k} alternative** precise, step-by-step computation graphs based on a provided database schema and data samples.

    ### OPERATORS ###
    You must strictly use ONLY the following atomic decomposition operations:
        --- I. Relational Operators ---
        1) SCAN:        "Return rows from [Table_Name]"
        2) FILTER:      "Return rows from [Previous_Step_ID] where [Column_Name_1] [Operator] [Value/Column_Name_2]"
        3) PROJECT:     "Return [Column_Names] of [Previous_Step_ID], calculating [Expression] if needed"
        4) AGGREGATE:   "Return [Agg_Func] of [Target_Column] grouped by [Grouping_Column] from [Previous_Step_ID]"
        5) SORT:        "Return [Previous_Step_ID] sorted by [Column_Name] [ASC/DESC]"
        6) LIMIT:       "Return the top [N] rows from [Previous_Step_ID]"
        7) JOIN:        "Return combined rows from [Previous_Step_ID_1] and [Previous_Step_ID_2] where [Join_Condition] matches"
        8) SET_OP:      "Return the [Union/Intersection/Difference] of [Previous_Step_ID_1] and [Previous_Step_ID_2]"
        9) DISTINCT:    "Return unique rows from [Previous_Step_ID] based on [Column_Names]"

        --- II. Semantic Operators ---
        10) LLM_DERIVE: "Return [Previous_Step_ID] with new column [New_Column_Name] derived from [Input_Columns] by [Instruction]"
        11) LLM_FILTER: "Return rows from [Previous_Step_ID] satisfying the semantic condition: [Instruction]"
        12) LLM_JOIN:   "Return combined rows from [Previous_Step_ID_1] and [Previous_Step_ID_2] using semantic matching logic: [Instruction]"
        13) LLM_AGGREGATE: "Return a summary of [Target_Column] grouped by [Grouping_Column] from [Previous_Step_ID] using instruction: [Instruction]"

        LEGEND:
        [Table_Name]: Exact name from Schema
        [Column_Name]: Exact column from Schema
        [Previous_Step_ID]: The 'id' of a step generated earlier
        [Agg_Func]: max, min, count, sum, avg
        [Operator]: !=, =, >, <, >=, <=, contains, in, not in, is null, is not null
        [Instruction]: Brief natural language description of the task, logic or condition.

    ### DIVERSITY STRATEGY ###
    Use the following principles to explore the solution space for generating plans:
    {diversification_strategy}

    ### GUIDELINES & CONSTRAINTS ###
    1) **Atomic Decomposition:** Each step must correspond to exactly one atomic operation from the list.
    2) **Schema Fidelity:** You must use the EXACT column and table names provided in the schema.
    3) **Value Inspection:** Do not rely solely on column names. Semantically cross-reference user terms with values in <data_preview>.
    4) **Dependency Graph:** The `parent` field must list the IDs of immediate predecessors.
    5) **Output Format:** Return ONLY a raw JSON object containing a list of plans.
    
    ### OUTPUT JSON SCHEMA ###
    {{
      "plans": [
        {{
            "steps": [{{"id": "step_2", "operator": "The operator name from the templates", "action": "The string description using the operator template", "parent": ["step_1"] }}]}}, ... (up to {k} plans)]
    }}
\end{lstlisting}
\end{tcolorbox}


\promptcaption{Planning Prompt}{Decomposition: User Prompt}
\begin{tcolorbox}[
    title=Decomposition: User Prompt,
    breakable, 
    sharp corners,
    colback=gray!3,
    colframe=gray!75,
    fonttitle=\bfseries,
    listing only,  
    listing options={
        basicstyle=\ttfamily\footnotesize,
        breaklines=true,
        columns=fullflexible,
        keepspaces=true
    }
]
    \footnotesize
\begin{lstlisting}
    ### INPUT ###

    ### DATA PREVIEW ###
    {data_preview}
    
    ### QUESTION ###
    {question}
    
    ### EXAMPLES ###
    | product_id | name        | description          | category |
    |------------|-------------|----------------------|----------|
    | A1         | iPhone 13   | Apple smartphone 5G  | Mobile   |
    | B2         | Galaxy S22  | Samsung phone        | Mobile   |

    User Question: "Show me the Apple phones."

    Response:
    {
        "plans": [
            {
                "steps": [
                    { "id": "step_1", "operator": "SCAN", "action": "Return rows from products", "parent": [products] },
                    { "id": "step_2", "operator": "FILTER", "action": "Return rows from step_1 where name contains 'Apple'", "parent": ["step_1"] }
                ]
            }
        ]
    }

\end{lstlisting}
\end{tcolorbox}

\promptcaption{Execution Prompt}{Step-to-SQL}
\begin{tcolorbox}[
    title=Step-to-SQL: Prompt,
    breakable, 
    sharp corners,
    colback=gray!3,
    colframe=gray!75,
    fonttitle=\bfseries,
    listing only,  
    listing options={
        basicstyle=\ttfamily\footnotesize,
        breaklines=true,
        columns=fullflexible,
        keepspaces=true
    }
]
    \footnotesize
\begin{lstlisting}
    You are an expert in text-to-SQL. Your task is to convert ONE atomic natural-language table step into a single SQLite-compatible SQL statement.
    
    ### CONSTRAINTS ###
    1) The available tables are: {tables}
    2) One step --> one SQL.
    3) If you create an output scalar or boolean, still return it as a SELECT ... so it forms a result table.
    4) Name any new column via AS.
    
    ### INPUT ###
    
    ### TABLE SCHEMAS ###
    {schema}
    
    ### TABLE PREVIEWS ###
    {preview_rows}
    
    ### ATOMIC STEP ###
    {step}
\end{lstlisting}
\end{tcolorbox}

\promptcaption{Execution Prompt}{Semantic Executor for FILTER, MAP, and AGGREGATE: System Prompt}
\begin{tcolorbox}[
    title= Semantic Executor for FILTER \& MAP \& AGGREGATE: System Prompt,
    breakable, 
    sharp corners,
    colback=gray!3,
    colframe=gray!75,
    fonttitle=\bfseries,
    listing only,  
    listing options={
        basicstyle=\ttfamily\footnotesize,
        breaklines=true,
        columns=fullflexible,
        keepspaces=true
    }
]
    \footnotesize
\begin{lstlisting}
You are an expert data-transformation and relational reasoning engine specialized in batch processing. Your responsibilities:
1) Execute semantic data transformations on structured tabular data
2) Return results in STRICT JSON format only
3) Preserve data integrity and handle edge cases gracefully

### CRITICAL OUTPUT FORMAT RULES ###
Your response must contain ONLY valid JSON with NO additional text, explanation, or Markdown.

For MAP and JOIN operations:
{
  "rows": [
    { "col1": value1, "col2": value2, ... },
    { "col1": value3, "col2": value4, ... }
  ]
}

For FILTER operations:
[0, 2, 5, ...]  // 0-based indices of matching rows

For AGGREGATE operations:
{
  "result": value_or_object,
}

### BEHAVIORAL GUIDELINES ###
- Process each row independently when specified
- Preserve original column values and data types
- Return ALL rows that match (for FILTER)
- Use semantic understanding (fuzzy matching, inference, etc.)
- Extract/derive values from existing columns only
\end{lstlisting}
\end{tcolorbox}

\promptcaption{Execution Prompt}{Semantic FILTER: User Prompt}
\begin{tcolorbox}[
    title=Semantic FILTER: User Prompt,
    breakable, 
    sharp corners,
    colback=gray!3,
    colframe=gray!75,
    fonttitle=\bfseries,
    listing only,  
    listing options={
        basicstyle=\ttfamily\footnotesize,
        breaklines=true,
        columns=fullflexible,
        keepspaces=true
    }
]
    \footnotesize
\begin{lstlisting}
### INSTRUCTION ###
{instruction}

### DATA ###
{data_str}

### TASK ###
Evaluate each row against the instruction. Return ONLY a JSON array of 0-based indices for rows that match. If no rows match, return an empty array.

### EXAMPLE ###
Input: Instruction: "Return items with price > 100"
       Data: [{{"id": 1, "price": 50}}, {{"id": 2, "price": 150}}, {{"id": 3, "price": 120}}]
Output: [1, 2]
\end{lstlisting}
\end{tcolorbox}

\promptcaption{Execution Prompt}{Semantic MAP: User Prompt}
\begin{tcolorbox}[
    title=Semantic MAP: User Prompt,
    breakable, 
    sharp corners,
    colback=gray!3,
    colframe=gray!75,
    fonttitle=\bfseries,
    listing only,  
    listing options={
        basicstyle=\ttfamily\footnotesize,
        breaklines=true,
        columns=fullflexible,
        keepspaces=true
    }
]
    \footnotesize
\begin{lstlisting}
### INSTRUCTION ###
{instruction}

### DATA ###
{data_str}

### TASK ###
Process each row independently. Apply the instruction to derive new columns or transform existing ones. Return the COMPLETE rows with all original columns PLUS any new derived columns.

### EXAMPLE ###
Input: Instruction: "Add column 'sentiment' by classifying tone as positive/negative"
       Data: [{{"id": 1, "text": "Great product!"}}, {{"id": 2, "text": "Terrible experience"}}]
Output: 
{{
  "rows": [
    {{"id": 1, "text": "Great product!", "sentiment": "positive"}},
    {{"id": 2, "text": "Terrible experience", "sentiment": "negative"}}
  ]
}}

### GUIDELINES ###
- Keep all original columns
- Add new columns as specified
- Maintain data types where possible
- Process row-by-row consistently
\end{lstlisting}
\end{tcolorbox}

\promptcaption{Execution Prompt}{Partial Semantic AGGREGATE: User Prompt}
\begin{tcolorbox}[
    title=Partial Semantic AGGREGATE: User Prompt,
    breakable, 
    sharp corners,
    colback=gray!3,
    colframe=gray!75,
    fonttitle=\bfseries,
    listing only,  
    listing options={
        basicstyle=\ttfamily\footnotesize,
        breaklines=true,
        columns=fullflexible,
        keepspaces=true
    }
]
    \footnotesize
\begin{lstlisting}
### INSTRUCTION ###
{instruction}

### DATA ###
{data_str}

### TASK ###
Summarize or aggregate this batch of data according to the instruction. This is a PARTIAL aggregation (may be merged with other partial results). Return a summary object that can be recursively combined with other partial results.

### EXAMPLE ###
Input: Instruction: "Calculate average salary"
       Data: [{{"name": "Alice", "salary": 80000}}, {{"name": "Bob", "salary": 90000}}]
Output: 
{{
  "sum": 170000,
  "count": 2,
  "average": 85000
}}
\end{lstlisting}
\end{tcolorbox}

\promptcaption{Execution Prompt}{Final Semantic AGGREGATE: User Prompt}
\begin{tcolorbox}[
    title=Final Semantic AGGREGATE: User Prompt,
    breakable, 
    sharp corners,
    colback=gray!3,
    colframe=gray!75,
    fonttitle=\bfseries,
    listing only,  
    listing options={
        basicstyle=\ttfamily\footnotesize,
        breaklines=true,
        columns=fullflexible,
        keepspaces=true
    }
]
    \footnotesize
\begin{lstlisting}
### INSTRUCTION ###
{instruction}

### DATA ###
{data_str}

### TASK ###
Perform final aggregation on the complete data (or merged partial results). Return a single result object matching the instruction.

### EXAMPLE ###
Input: Instruction: "Calculate average salary"
       Data: [{{"sum": 170000, "count": 2}}, {{"sum": 150000, "count": 2}}]
Output: 
{{
  "result": 80000,
  "summary": "Average salary across all employees"
}}
\end{lstlisting}
\end{tcolorbox}

\promptcaption{Execution Prompt}{Semantic Executor for JOIN: System Prompt}
\begin{tcolorbox}[
    title=Semantic Executor for JOIN: System Prompt,
    breakable, 
    sharp corners,
    colback=gray!3,
    colframe=gray!75,
    fonttitle=\bfseries,
    listing only,  
    listing options={
        basicstyle=\ttfamily\footnotesize,
        breaklines=true,
        columns=fullflexible,
        keepspaces=true
    }
]
    \footnotesize
\begin{lstlisting}
You are an expert data-transformation engine specializing in semantic JOIN operations.

### TASK ###
- Match rows from two tables based on semantic similarity or explicit join conditions
- Return ONLY the merged rows where matches are found
- Preserve all columns from both tables in the result

### OUTPUT ###
Return a JSON object:
{
  "rows": [
    { "col_a1": value, "col_b1": value, ... },
    { "col_a1": value, "col_b1": value, ... }
  ]
}

### GUIDELINES ###
 Include ALL columns from both tables in each result row
 Rename columns if needed to avoid conflicts (prefix with table name)
 Use semantic matching for fuzzy joins (e.g., "Google" matches "Alphabet Inc.")
 Return empty array if no matches found
\end{lstlisting}
\end{tcolorbox}

\promptcaption{Execution Prompt}{Semantic JOIN: User Prompt}
\begin{tcolorbox}[
    title=Semantic JOIN: User Prompt,
    breakable, 
    sharp corners,
    colback=gray!3,
    colframe=gray!75,
    fonttitle=\bfseries,
    listing only,  
    listing options={
        basicstyle=\ttfamily\footnotesize,
        breaklines=true,
        columns=fullflexible,
        keepspaces=true
    }
]
    \footnotesize
\begin{lstlisting}
### JOIN INSTRUCTION ###
{instruction}

### TABLE A ###
{table_a}

### TABLE B ###
{table_b}

### TASK ###
Find all matching pairs of rows (one from List A, one from List B) based on the instruction. For each match, combine the rows into a single object containing ALL columns from both tables.

### Example ###
Instruction: "Match users to orders by ID"
List A: [{{"user_id": 1, "name": "Alice"}}, {{"user_id": 2, "name": "Bob"}}]
List B: [{{"user_id": 1, "amount": 100}}, {{"user_id": 1, "amount": 50}}]
Output: 
{{
  "rows": [
    {{"user_id": 1, "name": "Alice", "amount": 100}},
    {{"user_id": 1, "name": "Alice", "amount": 50}}
  ]
}}
\end{lstlisting}
\end{tcolorbox}

\promptcaption{Evaluation Prompt}{Semantic Ground-truth Evaluator}
\begin{tcolorbox}[
    title=Semantic Ground-truth Evaluator: Prompt,
    breakable, 
    sharp corners,
    colback=gray!3,
    colframe=gray!75,
    fonttitle=\bfseries,
    listing only,  
    listing options={
        basicstyle=\ttfamily\footnotesize,
        breaklines=true,
        columns=fullflexible,
        keepspaces=true
    }
]
    \footnotesize
\begin{lstlisting}
You are an expert evaluator for database query results. Your task is to determine if a "Model Prediction" matches the "Ground Truth".

### RULES ###
1) Order Sensitivity: Treat the results as SETS. Row order does not matter unless the question explicitly asks for a ranking (e.g., "top 10").
2) Formatting: Ignore differences in formatting (e.g., "1,000" vs "1000", "$50" vs "50", "2023-01-01" vs "Jan 1, 2023").
3) Data Types: JSON objects, lists of tuples, and CSV strings should be compared based on content, not syntax.
4) Conversational Filler: If the prediction contains extra text (e.g., "The answer is 50"), extract the value "50" and compare it.
5) Column Names: Ignore column name differences (aliases) unless the user specifically asked for a specific column name.

### GROUND-TRUTH ###
{ground_truth}

### PREDICTION ###
{prediction}

### GUIDELINES ###
Think step-by-step:
1) Analyze the content of both the Ground Truth and Prediction.
2) Identify if there are differences in ordering, formatting, or wrapper text.
3) Determine if they convey the exact same data/information.

### OUTPUT ### 
- "reasoning": A brief string explaining why they match or differ.
- "verdict": "CORRECT" or "INCORRECT".
\end{lstlisting}
\end{tcolorbox}

\promptcaption{Consolidation Prompt}{Majority Vote}
\begin{tcolorbox}[
    title=Majority Vote: Prompt,
    breakable, 
    sharp corners,
    colback=gray!3,
    colframe=gray!75,
    fonttitle=\bfseries,
    listing only,  
    listing options={
        basicstyle=\ttfamily\footnotesize,
        breaklines=true,
        columns=fullflexible,
        keepspaces=true
    }
]
    \footnotesize
\begin{lstlisting}
You are an expert aggregator. I will provide you with a list of model-generated answers to a specific problem. Some answers might be phrased differently but mean the exact same thing.
    
### QUESTION ###
{question}

### PREDICTIONS ###
{prediction}

### TASK ###
1) Group the predictions that represent the same semantic answer/conclusion.
2) Identify which group has the most members (the plurality).
3) If a group has empty or unclear answers, ignore those.
4) If there is a tie for the largest group, choose any one of them.
5) Return ONLY the final answer from that majority group. No explanation.
\end{lstlisting}
\end{tcolorbox}

\promptcaption{Consolidation Prompt}{LLM-as-a-Judge Vote}
\begin{tcolorbox}[
    title=LLM-as-a-Judge Vote: Prompt,
    breakable, 
    sharp corners,
    colback=gray!3,
    colframe=gray!75,
    fonttitle=\bfseries,
    listing only,  
    listing options={
        basicstyle=\ttfamily\footnotesize,
        breaklines=true,
        columns=fullflexible,
        keepspaces=true
    }
]
    \footnotesize
\begin{lstlisting}
You are an expert judge evaluating different reasoning paths for the following question about a database. Your task is to select the plan that demonstrates the most logical, accurate, and complete reasoning:

### Task ###
Which Plan (index number) is the most likely to be correct? Respond ONLY with the integer index.


### QUESTION ###
{question}

### TABLES ###
{tables}

Below are several candidate plans and their resulting predictions. Select the plan that follows the most logical, accurate, and complete reasoning path.

### CANDIDATE PLANS ###
{plans}

### FEW-SHOT EXAMPLES ###
{few_shot_examples}
\end{lstlisting}
\end{tcolorbox}

\end{document}